%% file: main.tex
\pdfoutput=1
\documentclass[gmd, manuscript]{copernicus}

\graphicspath{{./figures/}} %Setting the graphicspath
\usepackage{enumitem}
\usepackage{hyperref}

\begin{document}
\nolinenumbers % needed for arXiv

\title{
Numerical coupling of aerosol
emissions, dry removal, and turbulent mixing 
in the E3SM Atmosphere Model version 1 (EAMv1), part I:
dust budget analyses and the impacts of a revised coupling scheme}

\Author[1]{Hui}{Wan}
\Author[1]{Kai}{Zhang}
\Author[2]{Christopher J.}{Vogl}
\Author[2]{Carol S.}{Woodward}
\Author[1]{Richard C.}{Easter}
\Author[3]{Philip J.}{Rasch}
\Author[4]{Yan}{Feng}
\Author[1]{Hailong}{Wang}

\affil[1]{Atmospheric Sciences and Global Change Division, Pacific Northwest National Laboratory, Richland, Washington, USA}
\affil[2]{Center for Applied Scientific Computing, Lawrence Livermore National Laboratory, Livermore, California, USA}
\affil[3]{Department of Atmospheric Sciences, University of Washington, Seattle, Washington, USA}
\affil[4]{Environmental Science Division, Argonne National Laboratory, Lemont, Illinois, USA}

\correspondence{Hui Wan (Hui.Wan@pnnl.gov)}
\runningtitle{Aerosol process coupling in EAMv1}
\runningauthor{Wan et al.}

\received{}
\pubdiscuss{} %% only important for two-stage journals
\revised{}
\accepted{}
\published{}
%% These dates will be inserted by Copernicus Publications during the typesetting process.

\firstpage{1}
\maketitle

\begin{abstract}
An earlier study evaluating the dust life cycle in the Energy Exascale Earth System Model (E3SM) Atmosphere Model version 1 (EAMv1) has revealed that the simulated global mean dust lifetime is substantially shorter when higher vertical resolution is used, primarily due to significant strengthening of dust dry removal in source regions.
This paper demonstrates that the sequential splitting of aerosol emissions,
dry removal, and turbulent mixing in the model's time integration loop, especially
the calculation of dry removal after surface emissions and before turbulent mixing, is the primary reason for the vertical resolution
sensitivity reported in that earlier study.
Based on this reasoning, we propose a simple revision to the numerical process coupling scheme, which
moves the application of the surface emissions to after dry removal and before 
turbulent mixing. The revised scheme allows newly emitted particles to be transported aloft 
by turbulence before being removed from the atmosphere,
and hence better resembles the dust life cycle in the real world.

Sensitivity experiments show that the revised process coupling substantially 
weakens dry removal and strengthens vertical mixing in dust source regions.
It also strengthens the large-scale transport from source to non-source regions,
strengthens dry removal outside the source regions, and strengthens wet removal 
and activation globally.
In wind-nudged simulations of the year 2010
with 1$^\circ$ horizontal grid spacing and 72 layers,
the revised process coupling leads to a 39\% increase in the global annual mean dust burden
and an increase of dust lifetime from 1.9 days to 2.6 days
when tuning parameters are kept unchanged.

The revised process coupling is implemented for all aerosol species in EAMv1.
The same qualitative changes in process rates are seen
in dust, sea salt, marine organic aerosols (MOA), black carbon (BC), and primary organic aerosols (POA),
as these species have significant sources from surface emissions.
Quantitatively, the changes are large for dust and sea salt
but are considerably smaller for the predominantly submicron species (i.e., MOA, BC, and POA).
The impacts on sulfate and secondary organic aerosols are
very small, as these species have little or no surface emissions.
\end{abstract}

\bigskip

\introduction  %% \introduction[modified heading if necessary]
\label{sec:introduction}

The numerical modeling of aerosol-climate interactions is a research topic with high levels of uncertainty \citep{Seinfeld_2016_PNAS,bellouin_2020_RFaer,smith_2020_RFaer}.
State-of-the-art global aerosol-climate models 
\citep[e.g.,][]{stier_2005_HAM1,mann_2010_glomap,Liu_et_al:2012,Zhang_et_al:2012,Wang_et_al:2020}
often consider a fairly large number of physical and chemical processes that affect aerosol life cycles.
Such processes include emissions, new particle formation, particle growth and aging,
transport by winds at local to global scales,
as well as removal from the atmosphere due to gravitational settling, boundary layer processes, 
formation of clouds and precipitation, and wet removal by precipitation.
Typically, the numerical representations of different processes are separately developed by subject matter experts 
and subsequently assembled to form global models. Different aerosol-climate models are known to show substantial discrepancies in the simulated magnitudes and spatiotemporal variations of the source and sink processes \citep{textor_2006_AeroCOM_I,gliss_2021_AeroCOM_III}.
% which, in turn, are known to significantly affect the simulated aerosol burden, spatial distribution, and lifetime.
Continuous efforts are being made to attribute such discrepancies to the physical assumptions
and computational methods used for individual processes, while
investigations on the impact of numerical process coupling \citep[e.g.,][]{wan:2013} are relatively rare.

The Energy Exascale Earth System Model (E3SM) is an Earth system model
developed by the U.S. Department of Energy for addressing
science questions related to the prediction of Earth system dynamics and climate change
\citep{E3SMv1_special_issue_intro,E3SMv1_code}.
The E3SM Atmosphere Model version 1 \citep[EAMv1,][]{Rasch_et_al:2019}
is an atmospheric general circulation model that includes 
a comprehensive representation of the life cycles of various natural and anthropogenic aerosol species.
A recent study by \citet{FengY_2022_JAMES_dust_in_EAMv1}
evaluated the dust life cycle in EAMv1 simulated with different horizontal and vertical resolutions.
It was shown that an EAMv1 simulation using 1$^\circ$ horizontal grid spacing and 72 vertical layers
produced a global mean dust lifetime of 1.85~days,
which was substantially shorter than the lifetime of 2.6~days reported by \citet{Liu_et_al:2012} and \citet{Scanza_2015_dust}
who used the predecessor model CAM5, the Community Atmosphere Model version 5 \citep{neale:2012},
with 2$^\circ$ horizontal grid spacing and 30 layers. 
\citet{FengY_2022_JAMES_dust_in_EAMv1} also showed that
reverting EAMv1's vertical resolution from 72 layers to 30 layers
would lengthen the global mean dust lifetime from 1.85~days to 2.4~days (see Table~4 therein),
primarily through the weakening of dry removal in the dust source regions.

This paper investigates the vertical resolution sensitivities
of dust lifetime and dry removal reported in \citet{FengY_2022_JAMES_dust_in_EAMv1}. 
Sections~\ref{sec:EAMv1_overview} and \ref{sec:doc_aerosols_in_EAMv1} 
provide a brief overview of EAMv1 and its parameterizations of aerosol emissions,
dry removal, and turbulent mixing.
Section~\ref{sec:coupling_scheme_in_default_EAMv1} describes 
EAMv1's numerical scheme used for coupling the aerosol-related processes.
Section~\ref{sec:motivation_for_revision} analyzes the simulated dust mass budget
and compares simulations conducted with 72 or 30 layers
to reveal weaknesses of the numerical process coupling in the publicly released EAMv1
(which we refer to as the original or default EAMv1 in this paper).
A simple revision to the numerical process coupling is proposed in Sect.~\ref{sec:revised_coupling}
and its impacts on the simulated aerosol climatology are evaluated
in Sect.~\ref{sec:impact_on_E3SM}.
The conclusions are drawn in Sect.~\ref{sec:conclusions}.

As is shown in Sect.~\ref{sec:impact_on_E3SM}, the revised process coupling
significantly weakens the dust dry removal in the source regions
and substantially reduces sensitivities of the simulated dust dry removal rate and
lifetime to vertical resolution. These changes appear to be desirable
given the deficiencies of the EAMv1 results pointed out in \citet{FengY_2022_JAMES_dust_in_EAMv1}.
On the other hand, the revision also results in 
large, global and systematical increases in the mass burden of dust and sea salt,
meaning the top-of-the-atmosphere energy fluxes become out of balance.
While such aerosol burden and energy flux changes can be offset by retuning
uncertain parameters used in the dust and sea salt emission parameterizations,
one might ask whether the revised coupling provides better results
(in terms of dry removal and dust lifetime) for the right reasons
and whether the retuning is worthwhile.
Other than presenting in Sect.~\ref{sec:motivation_for_revision} 
some process-level analyses of EAMv1 results to motivate the revision 
and confirming in Sect.~\ref{sec:impact_on_E3SM}
that the effects of the revised coupling on EAM's aerosol climatology
meet our expectation, it is not straightforward to
obtain additional evidence from EAM simulations to show that
the revised coupling is an improvement in the numerical sense.
This challenge has to do with the fact that the current EAM code does not allow for 
convergence testing of numerical process coupling between aerosol emissions, dry removal, and turbulent mixing 
without changing the timestep sizes and coupling frequencies of other physical processes in the model, for example resolved fluid dynamics and parameterized cloud processes.
To address this challenge,
the companion paper by \citet{Vogl_2023_dust_part_2}
presents a proof from an applied mathematics perspective
that the numerical error associated with dust dry removal 
caused by process splitting at each timestep is smaller 
when the revised scheme is used,
hence further justifying the adoption of the revised scheme in EAMv1.

%-----------------------------------------------------------------------------------
\input{2_model_description}

\input{3_motivation_for_revision}

\input{4_impact_on_E3SM}
\input{5_conclusions}
%-----------------------------------------------------------------------------------

\codedataavailability{
The EAMv1 source code used in this study can be found on Zenodo as record 
\href{https://doi.org/10.5281/zenodo.7995850}{7995850}
\citep{Wan_2023_Zenodo_7995850_code_v1_cflx_2021}.
The annual mean and instantaneous model output analyzed in this paper
can be found on Zenodo as records 
\href{https://doi.org/10.5281/zenodo.7996742}{7996742} \citep{Wan_2023_Zenodo_7996742_ann} and
\href{https://doi.org/10.5281/zenodo.8000745}{8000745} \citep{Wan_2023_Zenodo_8000745_inst}, respectively.
} %% use this section when having data sets and software code available

\clearpage

\appendix
\input{appendix_E3SM_figures}

\clearpage

\authorcontribution{
RCE and PJR initiated this study. 
HWan proposed and implemented the revised process coupling. 
KZ and HWan conducted and analyzed EAM simulations with input from the coauthors.
HWan wrote the manuscript with input from the coauthors. All coauthors contributed to the revisions.
} %% this section is mandatory

\competinginterests{The authors declare that no competing interests are present.} %% this section is mandatory even if you declare that no competing interests are present

\begin{acknowledgements}
This work was supported by the U.S. Department of Energy's (DOE's)
Scientific Discovery through Advanced Computing (SciDAC) program
via a partnership in Earth system modeling between DOE's
Biological and Environmental Research (BER) program and
Advanced Scientific Computing Research (ASCR) program.
KZ, YF, and HWang were supported by DOE BER through the E3SM project.
The work used resources of the National Energy Research Scientific Computing Center (NERSC),
a DOE Office of Science User Facility located at Lawrence Berkeley National Laboratory,
operated under Contract No. DE-AC02-05CH11231, using NERSC awards ASCR-ERCAP0025451.
Computational resources were also provided by the Compy supercomputer
operated for DOE BER by the Pacific Northwest National Laboratory (PNNL).
PNNL is operated for the U.S. DOE by Battelle Memorial Institute under contract DE-AC06-76RLO 1830.
The work by Lawrence Livermore National Laboratory was performed under the auspices of the U.S. Department of Energy under Contract DE-AC52-07NA27344.
LLNL-JRNL-850087-DRAFT.
YF acknowledges the support of Argonne National Laboratory (ANL) provided
by the U.S. DOE Office of Science, under Contract No. DE-AC02-06CH11357.
\end{acknowledgements}

\end{document}

%% file: 2_model_description.tex
%%%%%%%%%%%%%%%%%%%%%%%%%%%%%%%%%%%%%%%%%%%%%%%%%%%%%%%%%%%%
\section{The EAMv1 model and simulations}
\label{sec:model}
%%%%%%%%%%%%%%%%%%%%%%%%%%%%%%%%%%%%%%%%%%%%%%%%%%%%%%%%%%%%

The EAMv1 configuration described in \citet{Rasch_et_al:2019} and used in this study is a global hydrostatic atmospheric model that simulates the spatial distribution and time evolution of air temperature, pressure, winds, humidity, clouds, and precipitation. In addition, the model has 50 prognostic variables corresponding to the mixing ratios of aerosol particles of different sizes (diameters), chemical compositions, and attachment states
(interstitial or cloud-borne).
The mixing ratios of a few chemical gas species that are precursors of aerosols are also simulated using prognostic equations.

\subsection{EAMv1 overview}
\label{sec:EAMv1_overview}

EAMv1's dynamical core solves the primitive equations of the global atmospheric flow using a continuous Galerkin spectral-element method on a cubed-sphere horizontal mesh \citep{Dennis_et_al:2012,Taylor_Fournier:2010}. The vertical discretization uses a semi-Lagrangian scheme and a pressure-based terrain-following coordinate \citep{lin:2004}.
The resolved-scale tracer transport uses the discretization method of \citet{lin:2004} but has been adapted to the cubed sphere. The transport algorithm ensures local conservation of tracer and air masses as well as moist total energy \citep{Taylor_2011_bookchapter}.

For the parameterization of unresolved processes, the transfer of solar and terrestrial radiation is calculated with the Rapid Radiative Transfer Model for General Circulation Models \citep[RRTMG,][]{Iacono_et_al:2008,Mlawer_et_al:1997}. 
Deep convection is parameterized with the scheme of \citet{Zhang_McFarlane:1995} with modifications by \citet{Neale_et_al:2008} and \citet{Richter_Rasch:2008}.
Shallow convection, turbulence, and stratiform cloud macrophysics are represented by the higher-order closure parameterization named Cloud Layers Unified By Binormals \citep{larson_2017_tech_doc,Larson_and_Golaz:2005_MWR,Golaz_et_al:2002,Larson_et_al:2002}.
Stratiform cloud microphysics is represented with a two-moment parameterization with prognostic equations for the mass and number concentrations of cloud droplets, ice crystals, rain, and snow \citep{Gettelman_et_al:2015_MG2_part1,Gettelman_et_al:2015_MG2_part2,Morrison:2008}.
More detailed descriptions of the parameterization suite can be found in Section~2 of \citet{Rasch_et_al:2019} and the references therein.

\subsection{Aerosol processes in EAMv1}
\label{sec:doc_aerosols_in_EAMv1}

The Modal Aerosol Module 
\citep[MAM,][]{Wang_et_al:2020,Liu_et_al:2016,Liu_et_al:2012,Ghan_and_Easter_2006_cloud-borne_aerosol,Easter_MIRAGE_2004}
is a suite of parameterizations developed for global climate modeling,
aiming at providing a simplified yet sophisticated representation of aerosol life cycles
as well as their interactions with clouds and precipitation.
Seven aerosol species are considered in EAMv1:
sulfate, black carbon (BC), primary organic aerosols (POA), secondary organic aerosols (SOA),
marine organic aerosols (MOA), dust, and sea salt.
Mass and number concentration changes are predicted for aerosol particles of two attachment states,
interstitial and cloud-borne, which 
refer to the particles found outside and within cloud droplets, respectively.
The two attachment states correspond to two populations of aerosol particles. 
In the 4-mode configuration of MAM used in EAMv1 (i.e., MAM4),
the particle size distribution in each population is represented by 
one coarse mode and three fine-particle modes, see Fig.~2 in \citet{Wang_et_al:2020}.
Within a mode, the particle size distribution is represented by
a lognormal function of particle diameter, assuming the particles are spherical
and the geometric standard deviation of the lognormal function is fixed.
Under these assumptions, the mass mixing ratios of different species in different modes,
as well as the particle number mixing ratios of the different modes, are predicted. 

Aerosol processes currently considered in MAM4 include emissions
as well as particle mass and number changes resulting from
new particle formation (aerosol nucleation),
condensation and evaporation of chemical species, water uptake, coagulation,
aqueous chemistry in cloud droplets, 
aerosol activation (cloud nucleation),
resuspension from evaporating cloud droplets and precipitation,
in-cloud and below-cloud wet removal,
sub-grid vertical transport by deep convection and turbulence,
gravitational settling, and turbulent dry deposition.
Descriptions of MAM4 and its predecessors can be found in \citet{Rasch_et_al:2019}, \citet{Wang_et_al:2020}, \citet{Liu_et_al:2016}, \citet{Liu_et_al:2012}, \citet{Ghan_and_Easter_2006_cloud-borne_aerosol}, and \citet{Easter_MIRAGE_2004}.
Here we only briefly summarize the processes that are the foci of this study.

\begin{table*}[htbp]
\vspace{5mm}
\caption{\label{tab:emissions}
Global annual mean sources of aerosol mass (unit: Tg~yr$^{-1}$)
from surface emissions, elevated emissions, 
and production inside the atmosphere.
Also presented are the relative contributions (unit: \%) of these sources
to the total source of the corresponding species.
The "-" symbol means not applicable.
The results were derived from a wind-nudged simulation
of the year 2010 conducted with the default EAMv1 configuration
using 1$^\circ$ horizontal grid spacing and 72 layers
(simulation v1\_ori\_L72, see Sect.~\ref{sec:simulations}).
}
\begin{tabular}{lrrrrrrrrr}\tophline
\multirow{2}{*}{\bf Mass source of species} &
\multicolumn{2}{c}{\bf Surface emissions}  &&
\multicolumn{2}{c}{\bf Elevated emissions} &&
\multicolumn{2}{c}{\bf Atmospheric production} \\\cline{2-9}
&
Tg~yr$^{-1}$ & \% &&
Tg~yr$^{-1}$ & \% &&
Tg~yr$^{-1}$ & \% \\\middlehline
Dust                            & 3909.9 & 100\% &&    - & - &&  - & - \\ 
Sea salt                        & 3227.0 & 100\% &&    - & - &&  - & - \\ 
Marine organic aerosol (MOA)    &  21.95 & 100\% &&    - & - &&  - & - \\
Black carbon (BC)               &   7.76 &  80\% && 1.92 &  20\% &&  - & - \\ 
Primary organic aerosol (POA)   &   26.3 &  54\% && 22.3 &  46\% &&  - & - \\ 
Sulfate                         &   0.27 & 0.7\% && 1.50 & 3.6\% && 39.36 & 95.7\% \\ 
Secondary organic aerosol (SOA) &      - &     - &&    - &     - && 91.34 &  100\% \\ 
\bottomhline
\end{tabular}
\end{table*}

\begin{table*}[htbp]
\caption{\label{tab:emissions_by_mode}
Global annual mean surface emissions (unit: Tg~yr$^{-1}$) of 
selected aerosol species in different lognormal modes
and the range of geometric mean diameter $D_p$ of these modes.
Also presented are the relative contributions (unit: \%)
of the individual modes to the total surface emissions of the corresponding species.
The results were derived from a
wind-nudged simulation of the year 2010
conducted with the default EAMv1 configuration
using 1$^\circ$ horizontal grid spacing and 72 layers
(simulation v1\_ori\_L72, see Sect.~\ref{sec:simulations}). 
}
\scalebox{0.95}{
\begin{tabular}{lrrrrrrrrrrrrrr}\tophline
\multirow{3}{*}{\bf Surface emissions of species} &
\multicolumn{2}{c}{\bf Aitken mode}          &&
\multicolumn{2}{c}{\bf Primary carbon mode}  &&
\multicolumn{2}{c}{\bf Accumulation mode}    && 
\multicolumn{2}{c}{\bf Coarse mode} \\
&
\multicolumn{2}{c}{{\footnotesize   8.7~nm~$\leqslant D_p\leqslant $~52~nm}}     &&
\multicolumn{2}{c}{{\footnotesize    10~nm~$\leqslant D_p\leqslant $~100~nm}}    &&
\multicolumn{2}{c}{{\footnotesize  53.5~nm~$\leqslant D_p\leqslant $~440~nm}}    &&
\multicolumn{2}{c}{{\footnotesize 1~$\mu$m~$\leqslant D_p\leqslant $~4~$\mu$m}} \\\cline{2-12}
&
Tg~yr$^{-1}$ & \% &&
Tg~yr$^{-1}$ & \% &&
Tg~yr$^{-1}$ & \% &&
Tg~yr$^{-1}$ & \% \\\middlehline
Dust                            &   - &     - &&    - &       - &&  125.1 &  3\% &&  3784.8 & 97\% \\ 
Sea salt                        & 0.6 &   0\% &&    - &       - &&  106.3 &  3\% &&  3120.0 & 97\% \\ 
Marine organic aerosol (MOA)    & 0.1 &   1\% &&    - &       - &&   21.8 & 99\% &&       - &    - \\
Black carbon (BC)               &   - &     - &&  7.8 &   100\% &&      - &    - &&       - &    - \\
Primary organic aerosol (POA)   &   - &     - && 26.3 &   100\% &&      - &    - &&       - &    - \\ 
\bottomhline
\end{tabular}
}%scalebox
\end{table*}

\subsubsection{Emissions}
\label{sec:doc_emissions}

Since MAM considers a variety of sizes, species, and origins of aerosol particles,
a comprehensive set of assumptions and treatments are needed to specify aerosol emissions,
see Section~S1.1.1 in \citet[][]{Liu_et_al:2012}. 
In MAM4, the prescription of anthropogenic aerosol mass emissions follows protocols of 
model intercomparision projects and published studies.
The partitioning of the mass emissions into MAM's lognormal modes
and the calculation of aerosol number emissions in those modes are 
based on assumed emission size distributions.
Natural aerosol emissions are calculated online (i.e., during a simulation) 
using emission parameterizations. 
The scheme from \citet{Zender_2003_DEAD_model} is used for dust.
The calculation of sea salt emissions follows
\citet{Martensson_2003_JGR} for particle diameters from 20~nm to 2.5~$\mu$m 
and \citet{monahan_1986_SS} for particle diameters from 2.5~$\mu$m to 10~nm.
For MOA, the parameterization of \cite{Burrows_2014_OMF_parameterization}
is used to calculate the mass portion of MOA in the emitted sea spray aerosols;
this information is used in combination with the predicted sea salt emissions
to determine MOA emissions, assuming that 
MOA emissions add to the sea spray aerosol emissions, and that 
the emitted MOA is internally mixed with other aerosol species
\citep{Burrows_2022_MOA_in_EAMv1}.

Dust aerosols are assumed to be emitted only at the Earth's surface.
The emission fluxes are parameterized as a function of various properties 
of the Earth's surface
(e.g., soil moisture and erodibility)
and atmospheric conditions (e.g., friction velocity
and 10~m wind speed). Further details can be found in 
Section~2.4 of \citet{ZhangK_2016_GMD_subgrid_wind} and the references therein.
In mathematical terms, the emission of dust aerosols 
results in source terms in the evolution equations of dust mixing ratio
that are non-zero only at the bottom boundary of 
an atmosphere column.
While this is different from the treatment in the IMPACT global chemistry and transport model
which injects dust and some other aerosols uniformly in the boundary layer
\citep[see, e.g.,][]{Liu_2005_JGR,Wang_2008_JGR},
the assumption of non-zero dust emissions only at the Earth's surface
is commonly used in many global aerosol-climate models
\citep[e.g.,][]{Gong_2003_JGR,stier_2005_HAM1,mann_2010_glomap,ZhangK_2010_ACP_LIAM}.

Table~\ref{tab:emissions} shows the global annual mean magnitudes
of the surface and elevated emissions of the 7 aerosol species considered in MAM4.
To put the numbers into context, the mass sources 
caused by physical and chemical production inside the atmosphere 
are also shown, together with the percentage 
contribution of each source.
In Table~\ref{tab:emissions_by_mode}, the surface emissions of 
species that have no atmospheric production (i.e., only emissions as sources)
are broken into contributions from the 4 lognormal modes.
The numbers were derived from a
wind-nudged simulation of the year 2010
conducted with the default EAMv1 configuration
using 1$^\circ$ horizontal grid spacing and 72 layers. 
The simulation setup can be found in Sect.~\ref{sec:simulations}.

\subsubsection{Turbulent dry deposition and gravitational settling}
\label{sec:doc_dry_removal}

In this paper, we use the term ``dry removal" to refer to both
the gravitational settling and the turbulent dry deposition of aerosol particles,
as the two processes are handled together in the MAM4 parameterization suite embedded in EAMv1.
Gravitational settling is the downward movement of particles under the action of gravity,
whereas turbulent dry deposition refers to the loss of particles to the Earth's surface
through Brownian diffusion, impaction, interception, etc.
For both processes, the downward aerosol mass or number fluxes 
per unit time across a unit area
at the Earth's surface are calculated using the formula
\begin{equation}
M_{i,\text{sfc}} = \rho_{b}\, q_{i,b}\, v_{i,b}\,. 
\label{eq:dry_removal_flux_sfc}
\end{equation}
Here,
$M_{i,\text{sfc}}$ is the downward flux of the $i$-th aerosol tracer,
$\rho_b$ is air density in the bottom layer of the atmosphere
(i.e., the lowest model layer above the Earth's surface), and
$q_{i,b}$ is the mixing ratio of aerosol tracer $i$ in the bottom layer.
$v_{i,b}$ is the downward deposition velocity of aerosol tracer $i$
calculated using the aerosol properties and
ambient conditions in the bottom layer.

Like in many other models \citep[see, e.g.,][]{mann_2010_glomap,Zhang_et_al:2012},
MAM4 in EAMv1 assumes  
gravitational settling of aerosols can occur throughout the atmosphere column.
The resulting fluxes at altitudes above the Earth's surface
are parameterized as 
\begin{equation}
M_{i}^{\text{grav}}(z) = \rho(z)\, q_{i}(z)\, v^{\text{grav}}_{i}(z)\,, \label{eq:dry_removal_flux_sfc}
\end{equation}
where $z$ is geopotential height; 
$\rho(z)$, $q_{i}(z)$, $v^{\text{grav}}_{i}(z)$, and $M_{i}^{\text{grav}}(z)$
are air density, aerosol mixing ratio, gravitational settling velocity, 
and gravitational settling flux, respectively, at altitude $z$.
The calculation of the settling velocity  
is based on the Stokes' law, assuming particles reach their terminal velocities instantly.
The settling velocity of a single particle is calculated with Eqs.~(2) and (3) 
in \citet{Zhang_2001_aerosol_dry_deposition}. 
Correction factors are included to account for the impact of the 
lognormal size distribution.

The turbulent dry deposition velocity %, $v^\text{turb}_{i,b}$, 
is parameterized using Eq.~(21) in \citet{Zender_2003_DEAD_model}, 
for which the calculation of the quasi-laminar layer resistance follows 
Sect.~2 in \citet{Zhang_2001_aerosol_dry_deposition}.

\subsubsection{Turbulent mixing}
\label{sec:model_turbulent_mixing}

Turbulent mixing of aerosols in MAM4 is parameterized using the eddy diffusivity approach 
\citep[see, e.g.,][]{Garratt1994_ABL_book}, which gives aerosol concentration tendencies 
in the form of
\begin{equation}
\left(\dfrac{\partial \rho q_i}{\partial t}\right)^\text{turb-mix} = 
\dfrac{\partial}{\partial z}\left(\rho K_{h}\dfrac{\partial q_i}{\partial z}\right) \,,
\end{equation}
where $\rho$ is air density, $q_i$ is mixing ratio of aerosol tracer $i$,
$z$ is geopotential height, 
%$M^{\text{turb-mix}}_i$ is the mass flux of aerosol tracer $i$ at altitude $z$
%with the sign convention of positive upward, and
and $K_h$ is eddy diffusion coefficient.
In EAMv1, $K_h$ is calculated by the turbulence and cloud parameterization
CLUBB, while the turbulent mixing of aerosol mass, aerosol number, and cloud droplet number
is treated separately (outside CLUBB) in conjunction with aerosol
activation and resuspension from evaporating cloud droplets.
In other words, MAM4's parameterization of turbulent mixing and aerosol activation-resuspension 
solves differential equations in the form of 
\begin{equation}\label{eqn:dropmixnuc}
\left(\dfrac{\partial \rho q_i}{\partial t}\right)^\text{turb-mix+act/res} = 
\dfrac{\partial}{\partial z}\left(\rho K_{h}\dfrac{\partial q_i}{\partial z}\right) + 
\left(\dfrac{\partial \rho q_i}{\partial t}\right)^\text{act/res} \,,
\end{equation}
where the last term in Eq.~\eqref{eqn:dropmixnuc} is the 
rate of change caused by aerosol activation and resuspension from cloud droplets.
Additional information about the parameterization can be found in Section~S1.1.8
in the supplementary materials of \citet{Liu_et_al:2012} and in \citet{Ghan_and_Easter_2006_cloud-borne_aerosol}.

For clarification, we note that MAM also accounts for the resuspension of aerosols
from evaporating precipitation as a part of aerosol wet removal.
In this paper, when resuspension
is mentioned in conjunction with activation, we are referring to
the resuspension from cloud droplets.

\subsection{Numerical process coupling in EAMv1}
\label{sec:coupling_scheme_in_default_EAMv1}

\begin{figure*}[htbp]
\includegraphics[width=0.85\textwidth]{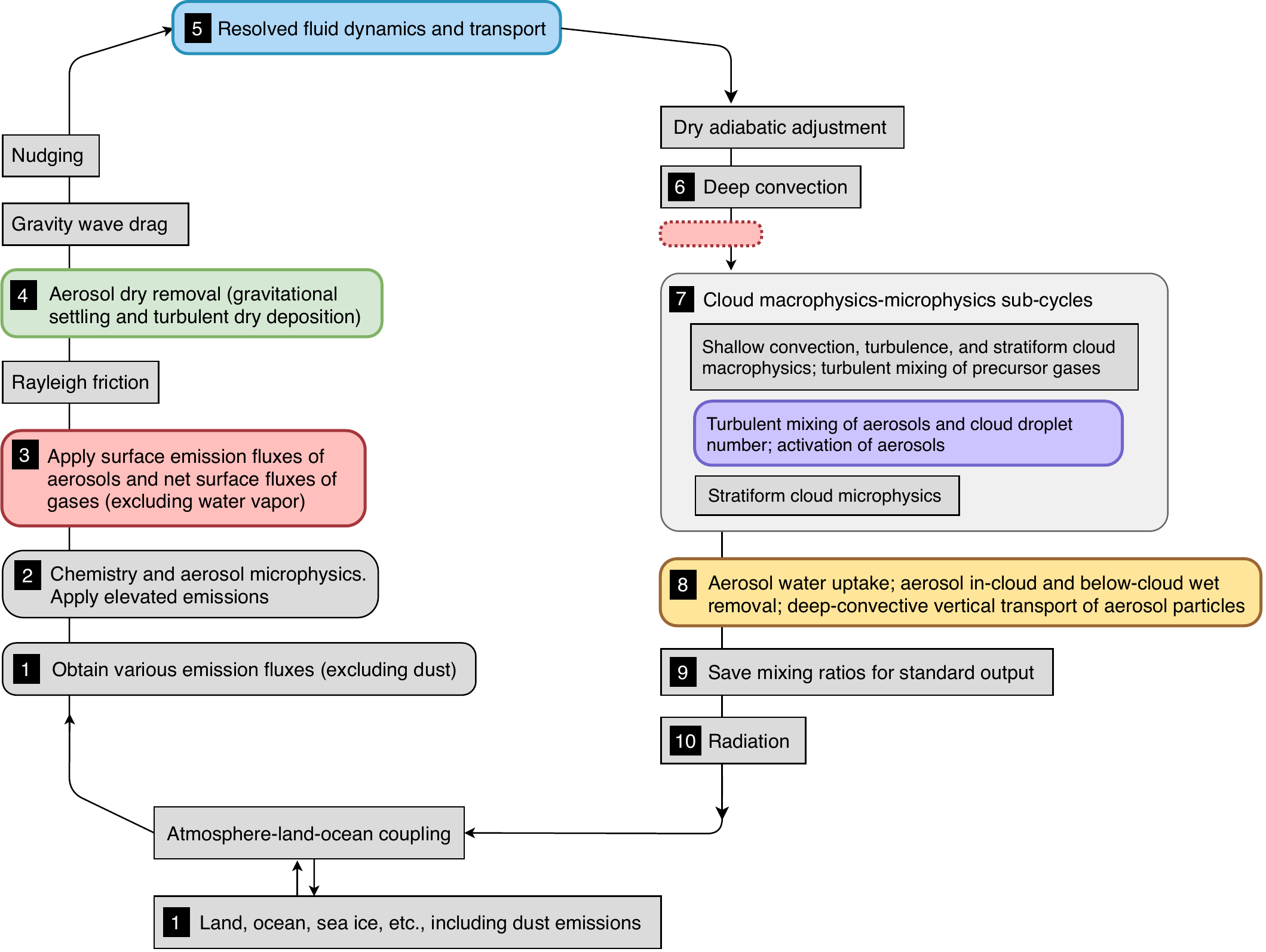}
\caption{\label{fig:flowchart_EAM}
A schematic showing the sequence of calculations in a 30-minute timestep in 1$^\circ$ EAMv1 simulations.
Rectangles are calculations that do not directly affect aerosol mixing ratios.
Boxes with round corners are calculations that change aerosol mixing ratios.
Colored boxes correspond to the physical processes for which
numerical results are shown in this paper.
The numbers in black squares correspond to the numbering
in Sect.~\ref{sec:coupling_scheme_in_default_EAMv1}.
The dashed pink box between deep convection (box 6) and 
the cloud sub-cycles (box 7) is where box 3 in the original EAMv1
is moved to when the revised coupling scheme is used.
}
\end{figure*}

The schematic in Fig.~\ref{fig:flowchart_EAM} depicts the sequence in which
the various atmospheric processes are calculated 
within a timestep of 30~minutes in the 1$^\circ$ EAMv1 simulations.
The schematic also shows where the coupling between EAM and the other components of E3SM (e.g., land and ocean)
occurs during the 30-minute timestep.
EAMv1 uses primarily the sequential splitting method for
the numerical coupling of aerosol-related processes and most other parameterizations.
With this method, the rates of change (i.e., tendencies)
of aerosol mixing ratios caused by a process (or a process group)
are calculated and then applied immediately to obtain new (updated) values of the mixing ratios.
The updated mixing ratios are then passed on to the next process or group
to calculate the next set of tendencies and update the mixing ratios.
Within a 30-minute timestep in EAMv1, the aerosol-related processes are calculated as follows.
(The numbering below matches the labels in Fig.~\ref{fig:flowchart_EAM}.)

\begin{enumerate}

\item Emission fluxes of aerosols are calculated based on atmospheric and Earth surface conditions 
or derived from emission datasets.
Dust emissions are calculated in E3SM's land model and passed to EAM by the coupler. 
Emissions of the other species are calculated (or read in from data files,
partitioned to MAM's lognormal modes,
and mapped to EAM's vertical grid) within the atmosphere model.

\item Gas-phase chemistry, aqueous-phase chemistry, and aerosol microphysics parameterizations
(gas-aerosol mass transfer, new particle formation, inter-mode transfer due to particle growth,
coagulation, and aging of primary carbon particles to accumulation mode)
are calculated for a 30-minute time window.
Elevated emissions of aerosols and precursor gases are also applied, as well as wet removal of the gases.
Rates of conversion between aerosol and gas species
and between different aerosol modes are calculated,
and the corresponding mixing ratio tendencies 
are used to update the aerosol and gas mixing ratios.

\item The surface fluxes of tracers (not including water vapor) stored 
in the Fortran variable \textsf{cam\_in\%cflx}
are converted to mixing ratio tendencies in 
the lowest model layer. These tendencies are applied over a 30-minute timestep
to update the corresponding mixing ratios.
We note that the surface fluxes applied at this location of the time loop
are the emission fluxes of aerosols and
the net fluxes (emission minus turbulent dry deposition) of precursor gases.

\item Dry removal
of aerosols is calculated for a 30-minute time window, 
and the aerosol mixing ratios are updated.
Gravitational settling affects all model layers where aerosols are present,
whereas turbulent dry deposition affects only the lowest model layer.

\item The large-scale transport scheme updates mixing ratios over two 15-minute vertical remapping timesteps,
each of which consists of three 5-minute sub-cycles of horizontal advection.

\item The deep convection parameterization changes EAM's temperature, humidity, and wind profiles as well as mixing ratios of hydrometeors but not yet the aerosol mixing ratios.

\item The parameterization of turbulent mixing and activation-resuspension of aerosol particles is calculated.
The cloud-borne and interstitial aerosol mass and number mixing ratios
are updated within this parameterization.
The tendencies of cloud droplet number mixing ratio are passed on to the
stratiform cloud microphysics parameterization.
In default 1$^\circ$ EAM simulations, the turbulent mixing and activation-resuspension
of aerosol particles, ice nucleation, 
the stratiform cloud microphysics, and the turbulence and shallow convection parameterization CLUBB are 
sub-cycled together using 5-minute timesteps 
(see Sect.~2.1 in \citeauthor{Wan_2021_GMD_time_step_sensitivity}, \citeyear{Wan_2021_GMD_time_step_sensitivity}
and Sect.~2 in \citeauthor{Santos_2021_time_step_sensitivity}, \citeyear{Santos_2021_time_step_sensitivity}).
Within each sub-cycle, CLUBB 
handles the turbulent transport of heat, water, and precursor gases;
CLUBB also 
provides the eddy diffusion coefficient, turbulent kinetic energy, and cloud fraction 
to the aerosol mixing and activation-resuspension parameterization.
The equation of turbulent mixing of aerosols is solved not by CLUBB but
in conjunction with aerosol activation-resuspension,
using an explicit time-stepping method with dynamically determined stepsizes.
%CLUBB itself and the ice nucleation and cloud microphysics parameterizations do not directly 
%affect aerosol mixing ratios in EAMv1.

\item After the turbulence and cloud microphysics sub-cycles, 
the processes of aerosol water uptake, 
aerosol in-cloud and below-cloud wet removal, and the vertical transport of aerosols
by deep convection are calculated for a timestep of 30 minutes, 
and the corresponding mixing ratios are updated. 

\item 
The mixing ratios, together with many other physical quantities describing
the simulated atmospheric state, are recorded and passed to the software infrastructure
that handles model output. In other words, 
the mixing ratios in EAM's output files that the model developers and users typically analyze
are from this location in the time loop.
(The mixing ratios at other locations presented later in the paper were obtained using
the online diagnostic tool CondiDiag \citep{Wan_2022_CondiDiag};
those additional mixing ratio values were included in the output files
under different variable names following the convention described in \citet{Wan_2022_CondiDiag}.)

\item The mixing ratios and other properties of aerosol particles are passed 
to the radiation parameterization where the aerosol impact on the atmospheric energy budget 
is calculated. Radiation does not directly affect aerosol mixing ratios.
\end{enumerate}
After radiation and some additional diagnostics, the atmosphere model exchanges
information with other components of E3SM, and the calculation goes back to step 1 
listed above.

The sequence of calculations described above is used by the original EAMv1.
In the revised process coupling scheme discussed in Sects.~\ref{sec:revised_coupling}
and \ref{sec:impact_on_E3SM}, 
the surface emissions of aerosols and the net surface fluxes (emissions minus turbulent dry deposition)
of precursor gases are applied after deep convection and before
the cloud macrophysics-microphysics sub-cycles.
That is, box 3 in Fig.~\ref{fig:flowchart_EAM}
is moved to the location indicated by the dashed pink box
(between boxes 6 and 7)
when the revised coupling scheme is used.

\subsection{Simulations}
\label{sec:simulations}

To analyze features of the simulated dust life cycle in the 1$^\circ$ configuration of EAMv1, 
we present in later sections 15-month simulations started from October 2009
using the default timestep settings documented in
\citet{Wan_2021_GMD_time_step_sensitivity} and \citet{Santos_2021_time_step_sensitivity}.
The results presented in this paper were derived from model output over the year 2010.

\begin{figure}[htbp]
\vspace{5mm}
 \includegraphics[height=0.45\textheight]{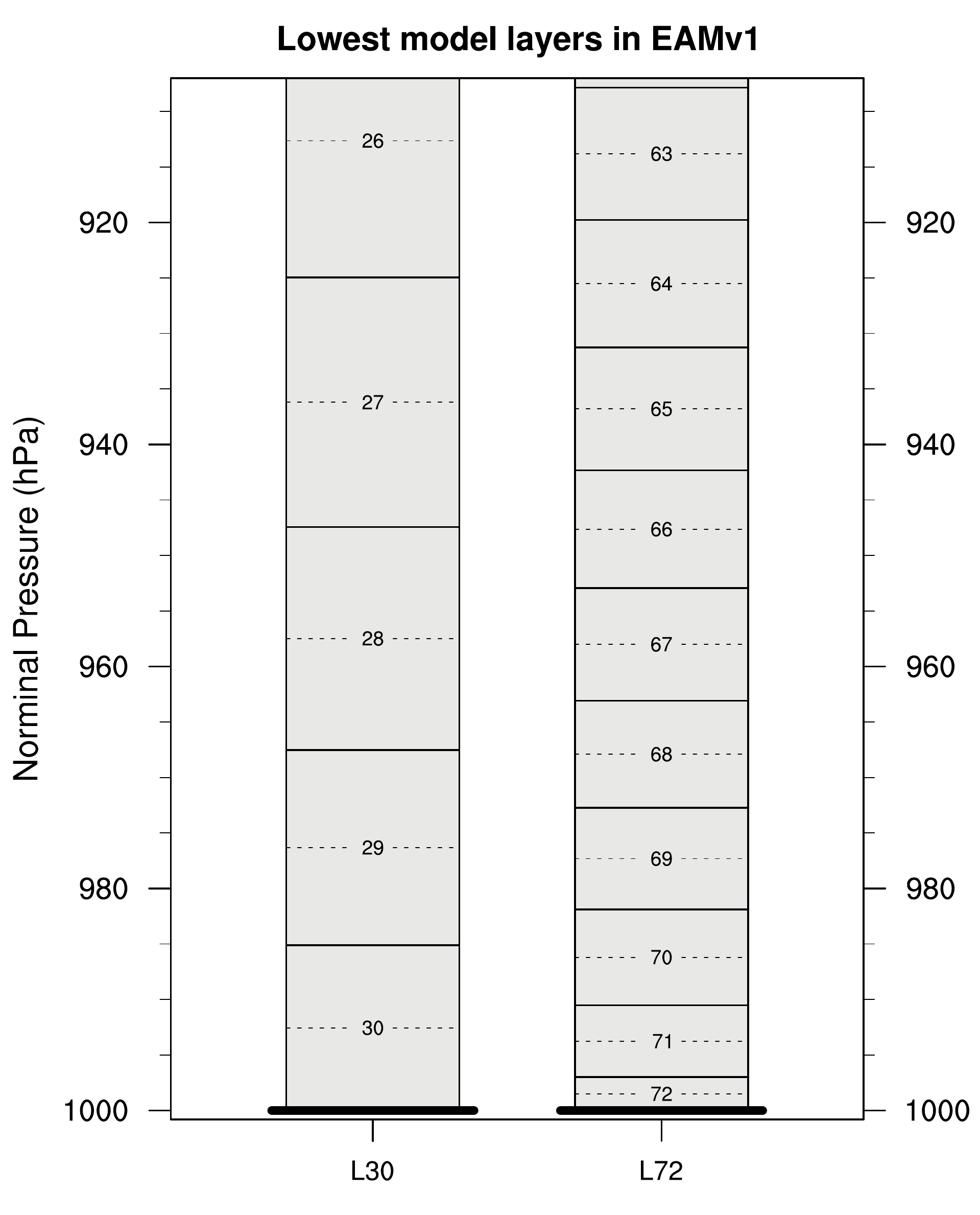}
 \caption{
  Near-surface layers of the L30 and L72 vertical grids used by EAMv1.
  Each gray box with a solid outline and a number at the center 
  represents one model layer. The numbers are layer indices.
  The dashed lines indicate locations of layer midpoints.
  The thick horizontal lines at 1000~hPa nominal pressure represent the Earth's surface.
  Here, nominal pressure refers to the air pressure at a layer interface or midpoint
  when the air pressure at the Earth's surface is 1000~hPa.}
  \label{fig:L72_L30_layers}
\end{figure}

A total of 4 simulations were conducted, two of which 
used EAMv1's original process coupling scheme described in Sect.~\ref{sec:coupling_scheme_in_default_EAMv1},
and the other two used the revised process coupling described in 
Sect.~\ref{sec:revised_coupling}.
For each of the coupling schemes, we conducted one simulation with 
EAMv1's default L72 vertical grid and another simulation with the L30 grid 
following the earlier studies of \citet{FengY_2022_JAMES_dust_in_EAMv1} and \citet{Zhang_et_al:2018}.
The simulation short names are listed in Table~\ref{tab:simulations}.
A comparison between the near-surface layers in the L30 and L72 grids
is shown in Fig.~\ref{fig:L72_L30_layers}.

The horizontal winds were nudged to 6-hourly 
ERA-Interim reanalysis 
\citep{Dee_2011_ERA-Interim} with a relaxation timescale of 6~hours following \citet{Sun_2019_JAMES}. 
In this study, wind nudging was applied to model levels
above 850~hPa approximately,
(i.e., above level 58 in the L72 grid and level 24 on the L30), 
as previous studies have shown that nudging winds near the Earth surface
can result in lower wind speeds compared to the free-running simulations,
and consequently affect the simulated dust emissions \citep{Timmreck_2004_JGR,Sun_2019_JAMES}.
The external forcings, e.g.,  
the sea surface temperature and sea ice extent as well as the 
emissions of anthropogenic aerosols and their precursors, 
followed the CMIP6 protocol of the twentieth-century transient simulations.
During a simulation, the evolution of dust mixing ratios within 30-minute timesteps 
and the tendencies associated with the colored boxes in Fig.~\ref{fig:flowchart_EAM}
were tracked with the online diagnostic tool of \citet{Wan_2022_CondiDiag}.

\begin{table}[htbp]
\vspace{5mm}
\caption{\label{tab:simulations}
Simulations conducted in this study. Further details can be found in Sect.~\ref{sec:simulations}.
}
\begin{tabular}{lcccccc}\tophline
{\bf Short name} & {\bf Process coupling scheme}   & {\bf Vertical grid} \\\middlehline
v1\_ori\_L72  & Original (Sect.~\ref{sec:coupling_scheme_in_default_EAMv1}) & L72 \\
v1\_ori\_L30  & Original (Sect.~\ref{sec:coupling_scheme_in_default_EAMv1}) & L30 \\
v1\_rev\_L72  & Revised (Sect.~\ref{sec:revised_coupling}) & L72 \\
v1\_rev\_L30  & Revised (Sect.~\ref{sec:revised_coupling}) & L30 \\
\bottomhline
\end{tabular}
\vspace{5mm}
\end{table}

%% file: 3_motivation_for_revision.tex
%------ time and space mean budget profiles--------------------
\begin{figure*}[htbp]
\includegraphics[width=0.98\textwidth]{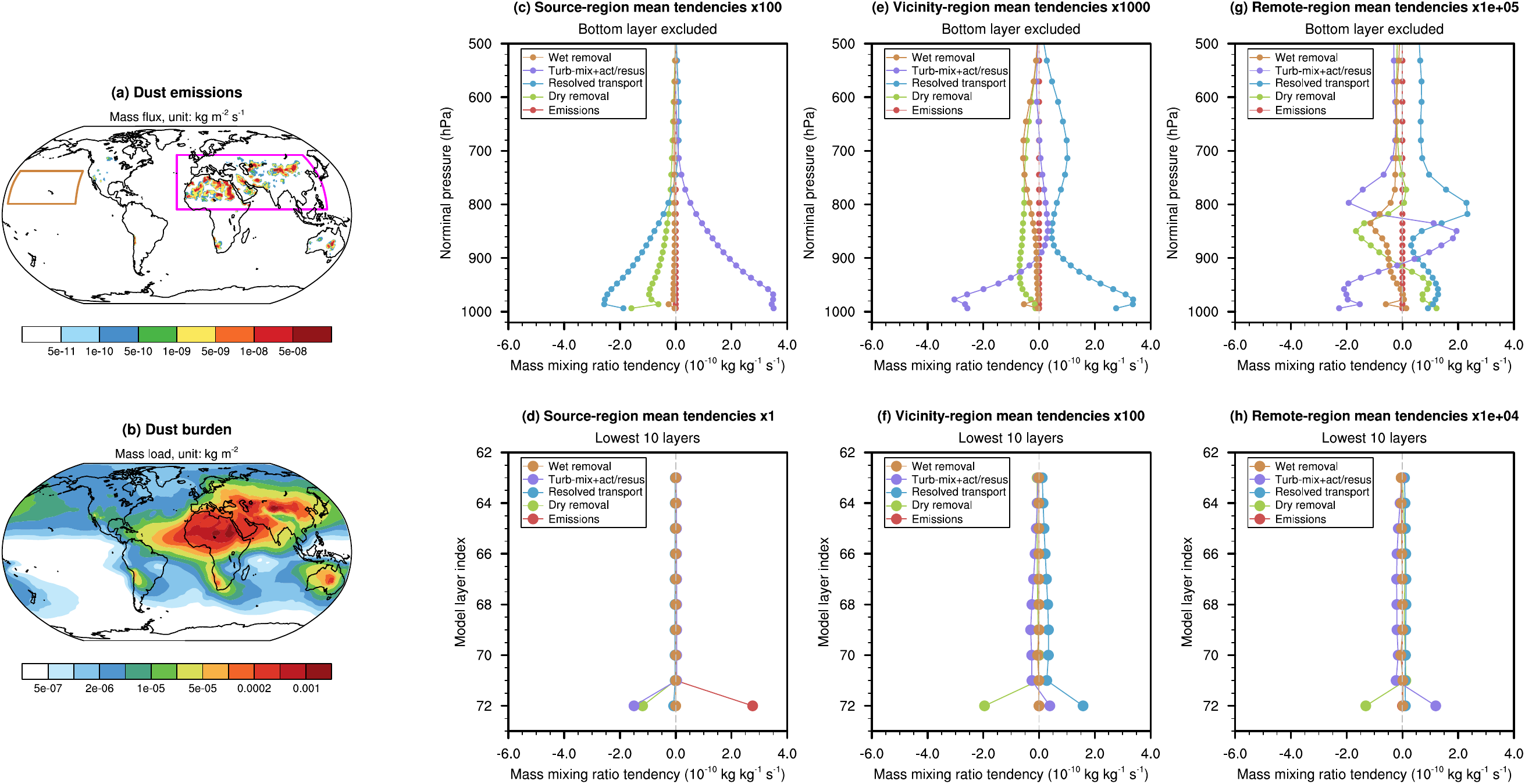}
\caption{\label{fig:dust_sources_and_sinks}
One-year averages of various quantities in simulation v1\_ori\_L72.
Left column: 
surface emission flux (panel a) and burden (panel b) of dust aerosol mass.
Other columns:
tendencies (i.e., rates of change) of interstitial dust aerosol mass mixing ratio caused by various physical processes, 
averaged over dust source regions (panels c, d), source-vicinity regions (panels e, f),
and remote regions (panels g, h).
Source regions and source-vicinity regions are defined as grid columns within the
magenta box in panel (a) in which the annual mean emission fluxes are non-zero and zero, respectively.
Remote regions refer to grid columns inside the brown box in panel (a).
Line plots in the upper row show results of the lower troposphere, 
with the bottom layer of the model's vertical grid (i.e., the grid layer closest to the Earth's surface) excluded.
Line plots in the lower row are results in the lowest 10 model layers including the bottom layer.
Note that the tendencies shown in panels (c) and (e)-(h) have been multiplied by 
factors of 100 to 100,000, as noted at the top of each panel, in order for 
the same x-axis range to be used for all profile plots.
}%\vspace{8mm}
\end{figure*}
%--------------

%%%%%%%%%%%%%%%%%%%%%%%%%%%%%%%%%%%%%%%%%%%%%%%%%%%%%%%%%%%%
\section{Motivation for a revised coupling scheme}
\label{sec:motivation_for_revision}
%%%%%%%%%%%%%%%%%%%%%%%%%%%%%%%%%%%%%%%%%%%%%%%%%%%%%%%%%%%%

%To reveal the weakness of the original coupling method used for the aerosol-related processes in EAMv1 and to explain the motivation for a revision, 
This section presents budget analyses for the total mass mixing ratio 
of interstitial dust particles (Sect.~\ref{sec:mass_budget})
and discusses the weaknesses of the original coupling scheme (Sect.~\ref{sec:original_coupling_weakness}).
A revised coupling scheme is described in Sect.~\ref{sec:revised_coupling}
and evaluated in the next section.

\subsection{Dust mass budget}
\label{sec:mass_budget}

Geographical distributions of the annual mean dust mass emission fluxes 
and interstitial dust burden from the default EAMv1 configuration 
(i.e., simulation v1\_ori\_L72) are presented in panels (a) and (b) of Fig.~\ref{fig:dust_sources_and_sinks}.
The mass burden shown here is the total mass mixing ratio summed over MAM4's 
accumulation mode and coarse mode, 
multiplied by air density, and vertically integrated over the atmosphere column. 
As expected, the dust emissions are highly inhomogeneous in space
while the distribution of the dust burden is much smoother due to transport by winds.

To get an overview of the balance between different physical processes, 
we selected the magenta box in Fig.~\ref{fig:dust_sources_and_sinks}a to cover
the major dust sources in Asia, Europe, and North Africa.
Within the magenta box, the grid columns with 
non-zero annual mean emissions were marked as the source regions and 
the grid columns with no emissions were marked as the source-vicinity regions. 
The brown box in Fig.~\ref{fig:dust_sources_and_sinks}a 
was selected to represent regions far away from dust emissions (i.e., the remote regions).
The annual mean dust mass mixing ratio tendency profiles were 
averaged over the source, vicinity, and remote regions 
and are shown in Fig.~\ref{fig:dust_sources_and_sinks}c-h.

The characteristic tendency magnitudes 
in the three types of regions as well as at different altitudes can be seen in 
Fig.~\ref{fig:dust_sources_and_sinks}c-h. The upper row 
is the results in the lower troposphere with the lowest model layer {\it excluded},
while the lower row shows the 10 model layers closest to the Earth surface, 
{\it including} the bottom layer.
Furthermore, the tendencies in panels c and e-h have been 
multiplied by factors of 100 to 100,000
in order to be plotted with the same x-axis range as in panel d.
These factors of multiplication are noted at the top of each panel.
A key feature revealed by these vertical profiles is that the annual and regional mean tendencies 
in the lowest model layer in the dust source regions are 
2 orders of magnitude stronger than the tendencies in the upper layers within the same regions.
Furthermore, the tendencies in the lowest model layer in the source regions 
are 2-4 orders stronger than the tendencies in any layer in the source-vicinity or remote regions.
The dominant source and sink terms in the global-scale annual mean dust mass budget are the 
following processes in the lowest model layer in the dust source regions:
(1) surface emissions, (2) dry removal, and (3) turbulent mixing and aerosol activation-resuspension.

\begin{figure}[htbp]
\includegraphics[width=0.45\textwidth]{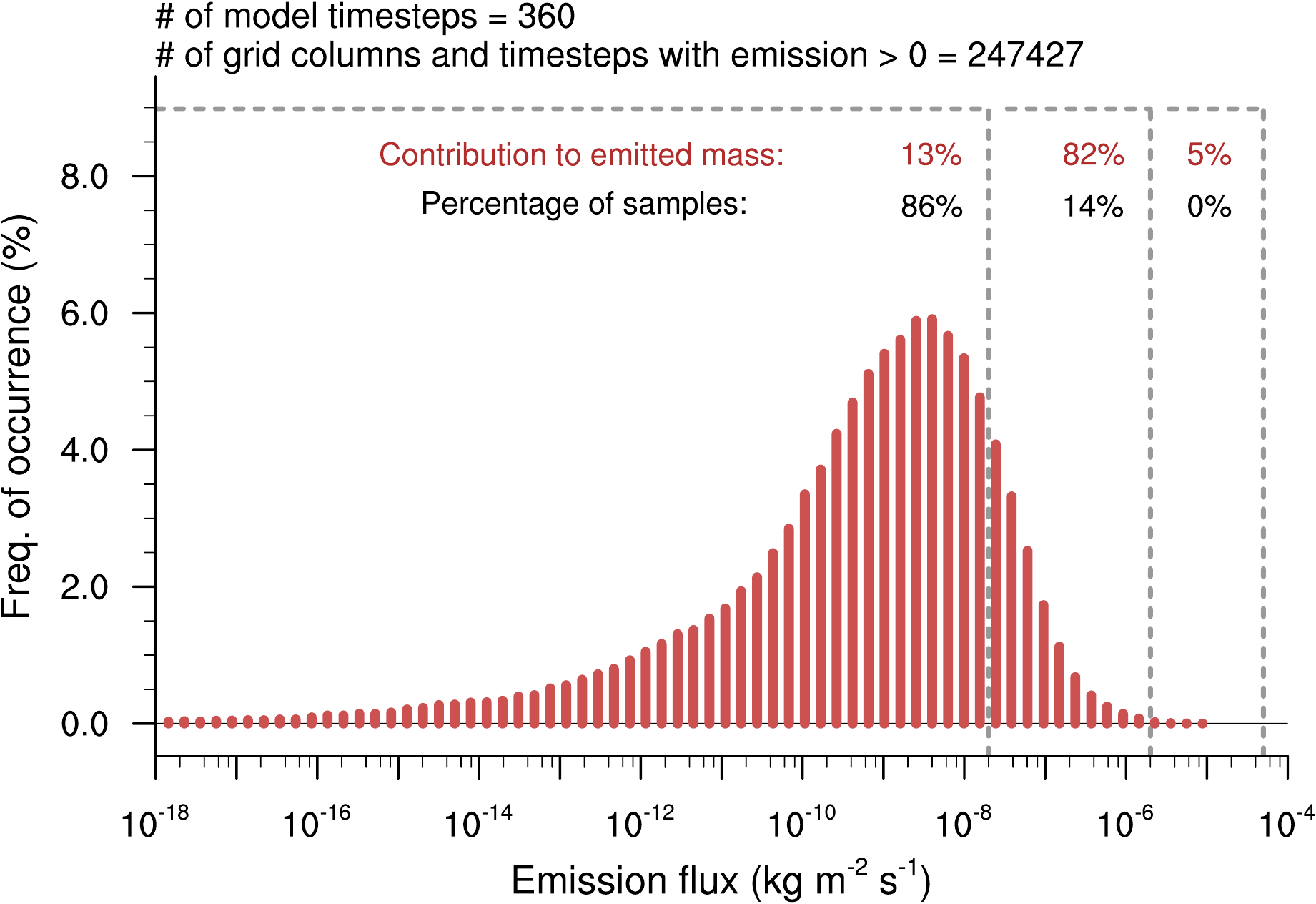}
\caption{\label{fig:emission_pdf}
Histogram of dust emission fluxes derived from 6-hourly instantaneous model output
in the 90-day period of 2010-01-29 to 2010-04-28
in grid columns with non-zero dust emissions in the magenta box marked in 
Fig.~\ref{fig:dust_sources_and_sinks}a.}
\end{figure}

\begin{figure*}[htbp]
\includegraphics[width=\textwidth]{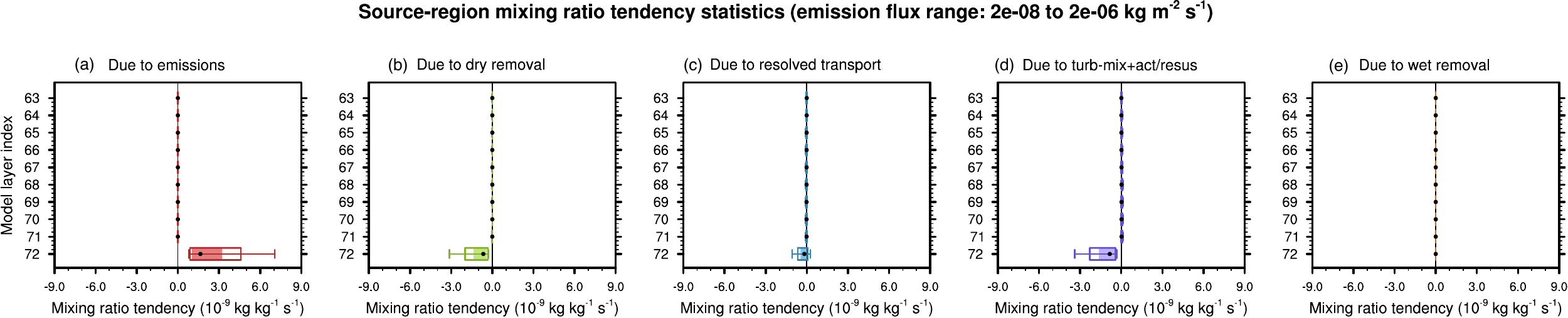}
\caption{
Statistical distributions of dust mixing ratio tendencies 
in the lowest 10 layers in the default EAMv1 (i.e., simulation v1\_ori\_L72)
derived from grid columns and timesteps
corresponding to the middle portion of the 
emission flux histogram in Fig.~\ref{fig:emission_pdf}.
Different panels show results for different physical processes.
Within each panel and for each model layer, 
the filled box depicts the middle 50\% of the statistical distribution
of the instantaneous mixing ratio tendencies;
the unfilled box corresponds to the middle 67\% of the distribution;
the vertical whiskers are the 10th and 90th percentiles;
the black dot indicates the median.
Results for the other two portions of the emission histogram shown in Fig.~\ref{fig:emission_pdf},
corresponding to dust source regions with instantaneous emission fluxes
weaker than 2$\times$10$^{-8}$~kg~m$^{-2}$~s$^{-1}$ or stronger than 
2$\times$10$^{-6}$~kg~m$^{-2}$~s$^{-1}$, 
are shown in Fig.~\ref{fig:dust_tend_box_plot_source_2}.
Results for the source-vicinity and remote regions are shown 
in Fig.~\ref{fig:dust_tend_box_plot_vicinity_and_remote}.
}
\label{fig:dust_tend_box_plot_source}
\end{figure*}

To demonstrate that similar contrasts in the magnitudes of process rates 
can be seen in individual grid columns and at individual timesteps,
Fig.~\ref{fig:emission_pdf} and Fig.~\ref{fig:dust_tend_box_plot_source} present results 
derived from 6-hourly instantaneous model output in the 90-day period 
of 2010-01-29 to 2010-04-28.
Figure~\ref{fig:emission_pdf} is a histogram of the dust mass emission fluxes in the magenta box 
in Fig.~\ref{fig:dust_sources_and_sinks}a,
derived from grid columns and timesteps with non-zero dust emissions. 
The instantaneous emission fluxes span more than 10 orders of magnitude. 
About 82\% of the emitted mass is attributable to events with mass fluxes 
between 2$\times$10$^{-8}$~kg~m$^{-2}$~s$^{-1}$ and 2$\times$10$^{-6}$~kg~m$^{-2}$~s$^{-1}$;
about 13\% of the emitted mass is contributed by a large number of weak emission events 
and the remaining 5\% by a very small number of very strong events (Fig.~\ref{fig:emission_pdf}).

For the middle portion of the histogram, 
Fig.~\ref{fig:dust_tend_box_plot_source} shows the statistical distributions of 
the instantaneous dust mixing ratio tendencies in the 
lowest 10 model layers of the L72 grid, where the 
different panels correspond to different physical processes.
The results for the other two portions of the histogram as well as 
for the source-vicinity and remote regions are shown in 
Figs.~\ref{fig:dust_tend_box_plot_source_2} and
 \ref{fig:dust_tend_box_plot_vicinity_and_remote}, respectively.
These figures suggest the dominant sources and sinks of dust mass 
on a timestep-by-timestep and grid-column-by-grid-column basis
are the same as what we saw in the regional and annual averages, namely the 
following processes in the lowest model layer in the dust source regions:
(1) surface emissions, (2) dry removal, and (3) turbulent mixing and aerosol activation-resuspension.

\vspace{5mm}
%------budget figure--------------------
%--------------

\begin{figure*}[htbp]
\includegraphics[width=\textwidth]{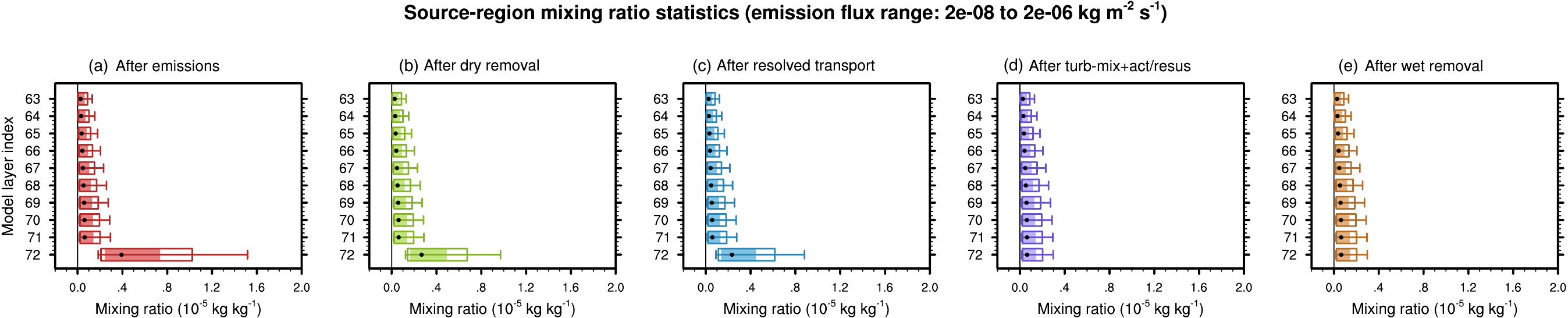}
\caption{
Similar to Fig.~\ref{fig:dust_tend_box_plot_source} but showing the statistical distributions
of instantaneous dust mass mixing ratios after the tendencies shown in 
Fig.~\ref{fig:dust_tend_box_plot_source} have been applied.
Results here correspond to the middle portion of the emission histogram in 
Fig.~\ref{fig:emission_pdf}.
Results for the source-vicinity and remote regions are shown in 
Fig.~\ref{fig:dust_state_box_plot_vicinity_and_remote}.
}
\label{fig:dust_state_box_plot_source}
\end{figure*}

%---------------------------------------------
\subsection{Weaknesses of the original coupling scheme}
\label{sec:original_coupling_weakness}
%---------------------------------------------

Since sequential splitting is used in the default EAMv1 for the various aerosol processes
discussed above, and the ordering is emissions, dry removal, resolved transport, 
turbulent-mixing-activation-resuspension, and wet removal,
we show in Fig.~\ref{fig:dust_state_box_plot_source} 
the statistical distributions of instantaneous dust mixing ratios
after each of these processes has been calculated and
the tendencies have been applied (i.e., the mixing ratios have been updated).
The results in this figure (Fig.~\ref{fig:dust_state_box_plot_source})
correspond to the same grid columns and timesteps 
as shown in Fig.~\ref{fig:dust_tend_box_plot_source}.
Given the nature of the sequential splitting method and 
the magnitudes of tendencies seen in Fig.~\ref{fig:dust_tend_box_plot_source},
it is understandable that large mixing ratio spikes are seen in the lowest model layer 
after the surface emissions are applied (Fig.~\ref{fig:dust_state_box_plot_source}a). 
The mixing ratio profiles in Fig.~\ref{fig:dust_state_box_plot_source}
also indicate that although dry removal is a strong sink in the lowest model layer
and resolved transport can be a significant sink, 
neither is sufficiently strong to remove the near-surface mixing ratio spikes.
In contrast, the turbulent mixing and aerosol activation--resuspension 
processes are very effective in vertically smoothing the profiles.
Since the dust source regions are typically dry and the lowest model layer is thin
(about 20~m on average), we do not expect aerosol activation to occur there frequently;
hence the smoothing effect is most likely attributable to turbulent mixing.
Using the online diagnostic tool of \citet{Wan_2022_CondiDiag}, 
we also analyzed the dust mixing ratios after each of the 
5-minute sub-cycles used for the parameterizations of
turbulence mixing and aerosol activation-resuspension.
We found that the spikes of dust mixing ratio were typically eliminated 
after 1 or 2 sub-cycles (i.e., 5 to 10 minutes). 
This is consistent with the results from \citet{Wang_Rasch_Feingold_2011_Sc_engineering},
who showed that particles injected from ships traveling below marine stratocumulus
can be lofted and mixed through the cloud-topped boundary layer 
within minutes \citep[see Fig.~1c and Sect.~3.1 in][]{Wang_Rasch_Feingold_2011_Sc_engineering}.

While the sub-timestep evolution of dust mixing ratio shown in 
Fig.~\ref{fig:dust_state_box_plot_source} is to be expected from 
the use of sequential splitting and
the characteristic magnitudes of process rates in the default EAMv1,
the figure provides a clue that the ordering of the various process
has severe deficiencies.
In the real world, dust emissions typically occur in turbulent atmospheric environments
where the turbulent eddies are efficient in transporting the emitted particles aloft.
In EAM, when the emission fluxes are applied  
to update the dust mixing ratios in the lowest model layer, 
the effect can be interpreted as immediately mixing 
the emitted particles throughout the lowest model layer
but also temporarily trapping the particles within that layer, 
resulting in high concentrations being passed to the next process in the time loop,
which is dry removal in the default EAMv1.
Since the dry removal fluxes at the Earth's surface 
are proportional to the mixing ratios in the lowest model layer (see Sect.~\ref{sec:doc_dry_removal},
Eq.~\ref{eq:dry_removal_flux_sfc}),
the high mixing ratios depicted in Fig.~\ref{fig:dust_state_box_plot_source}a
are expected to lead to strong dry removal fluxes.
If one chose, instead, to calculate turbulent mixing immediately after the emissions are applied,
and then calculate dry removal afterwards (i.e., after turbulent mixing),
then the input to the dry removal parameterization
would look similar to Fig.~\ref{fig:dust_state_box_plot_source}d (i.e., without a spike in the bottom layer),
hence resulting in mush weaker dry removal fluxes at the surface.
Following this logic, we expect the EAM simulations to be
sensitive to the ordering of the emission, turbulent mixing, and dry removal processes.

\begin{figure}[htbp]
 \includegraphics[width=0.38\textwidth]{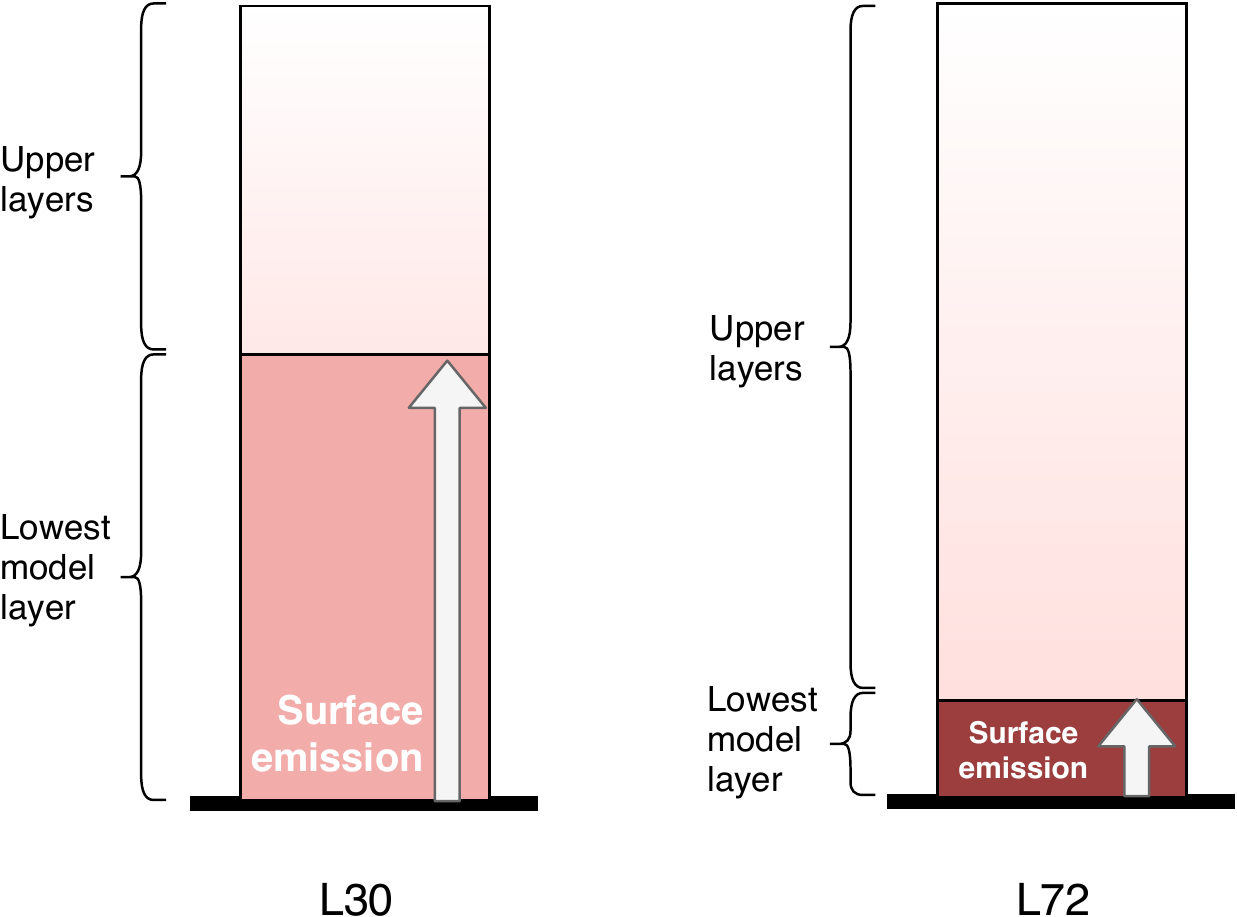}
 \caption{
  A schematic depicting the impact of layer thickness on the dust mixing ratio
  in the lowest model layer after surface emissions are applied.
  In a numerical model where the surface emissions are treated as a separate physical process
  and coupled with other processes using sequential splitting,
  given the same emission flux and same timestep size,
  a thinner bottom layer means a higher layer-mean mixing ratio 
  will result after the surface emissions are applied.
  The left and right parts of the schematic correspond to the L30 and L72 vertical grids 
  depicted in Fig.~\ref{fig:L72_L30_layers}, respectively.
  A discussion of this schematic can be found in Sect.~\ref{sec:original_coupling_weakness}.
  }
  \label{fig:bottom_layer_thickness_and_emission}
\end{figure}

\begin{figure*}[htbp]
\includegraphics[width=\textwidth]{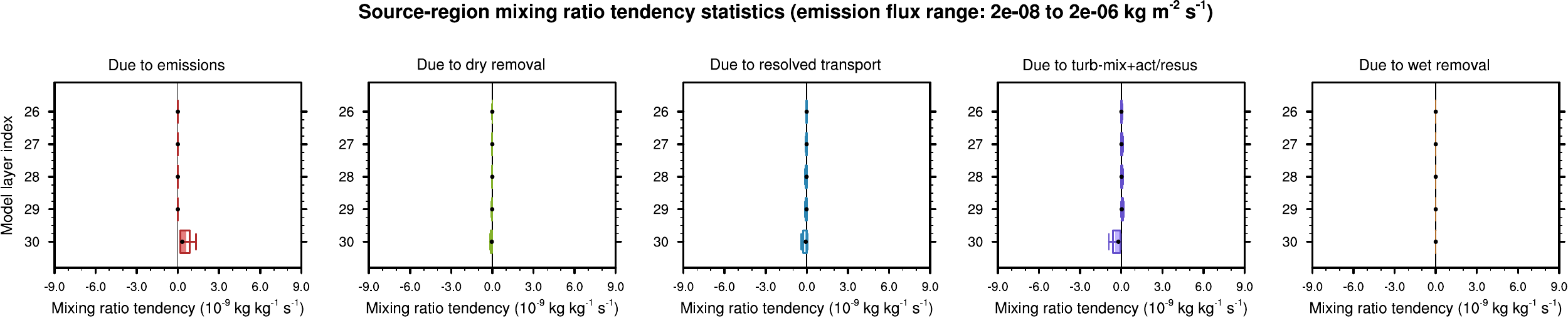}\vspace{4mm}
\includegraphics[width=\textwidth]{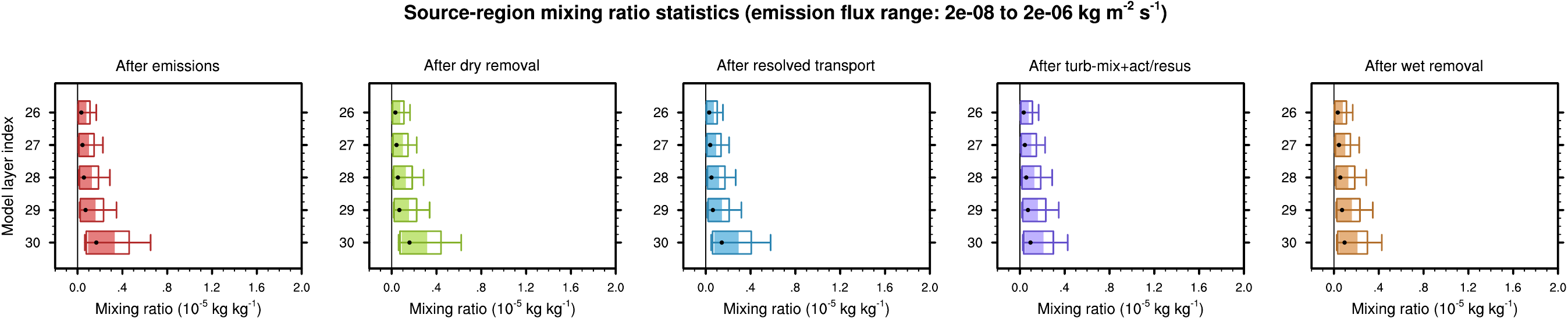}
\caption{
Similar to Figs.~\ref{fig:dust_tend_box_plot_source} and \ref{fig:dust_state_box_plot_source}
but showing results from simulation v1\_ori\_L30 which used the L30 grid instead of L72.
The upper row shows statistical distributions of the dust mass mixing ratio tendencies
caused by different aerosol processes.
The lower row shows the statistical distributions of dust mass mixing ratios 
after the corresponding tendencies in the upper row have been applied.
Only 5 model layers are shown in this figure, as Fig.~\ref{fig:L72_L30_layers}
suggests that the 5 lowest layers in the L30 grid cover roughly the same altitude range
as the 10 lowest layers in the L72 grid.
}
\label{fig:dust_tend_and_state_box_plot_source_L30}
\end{figure*}

The reasoning above also provides an explanation for the strong 
vertical resolution sensitivity of dust dry removal reported in \citet{FengY_2022_JAMES_dust_in_EAMv1}.
As mentioned above, dust is emitted only into the lowest model layer
in EAMv1, meaning the particles are temporarily trapped below the upper interface of the layer.
The lowest layer in the L72 grid 
is about 1/5 in thickness compared to the lowest layer in the L30 grid
(Fig.~\ref{fig:L72_L30_layers}).
Therefore, given the same emission fluxes, the temporary increases 
in dust mixing ratio after the emissions are applied in a simulation using the L72 grid 
are expected to be 5 times as high as in another simulation that uses the L30 grid
(see schematic in Fig.~\ref{fig:bottom_layer_thickness_and_emission}), 
which in turn can lead to significantly stronger dry deposition in the L72 simulation.
This expectation is confirmed by Fig.~\ref{fig:dust_tend_and_state_box_plot_source_L30}
which shows results from simulation v1\_ori\_L30.
In the L30 simulation, both the tendencies and the mixing ratios spikes in the lowest model layer 
are substantially weaker than in v1\_ori\_L72.

It is worth clarifying that the process coupling issue identified here is not specific to EAMv1. In the predecessor model CAM5, although the atmospheric turbulence parameterization used the scheme of \citet{Park_Bretherton:2009} and was calculated before aerosol dry removal and after surface emissions, the aerosol tracers were {\it not} mixed by the \citet{Park_Bretherton:2009} parameterization, but rather by the same turbulent mixing and  aerosol activation--resuspension parameterization as in EAMv1. In other words, as far as the aerosol tracers are concerned, the sequence of calculation in CAM5 was the same as in EAMv1. We expect that if CAM5 simulations are performed with EAM's L72 grid, significantly stronger dry removal and shorter dust lifetime will result as well.

Furthermore, we note that the same process coupling issue also exists for other aerosol species in EAM
that have surface emissions, although the magnitude of the impact depends on
the relative importance of the surface emissions as well as the 
typical sizes of the particles. 
This is further discussed in Sect.~\ref{sec:EAM_results_other_species}
using EAM results.

The precursor gases in EAMv1, on the other hand, do not suffer from this coupling issue, 
because the splitting and ordering of the gas-related processes are different.
Precursor gases in the model are assumed to experience turbulent dry deposition but
no gravitational settling. The surface dry removal fluxes of gases are 
calculated after gas-phase chemistry inside the box marked with ``2'' in the 
schematic in Fig.~\ref{fig:flowchart_EAM} and then, instead of being used to update gas mixing ratios,
the dry removal fluxes are subtracted from the surface emission fluxes,
and the residual (i.e., the net flux) is saved in the Fortran variable {\sf cam\_in\%cflx}.
When {\sf cam\_in\%cflx} is used to update tracer mixing ratios in the lowest model layer
in box 3 in the schematic, it is the residual of emission and dry deposition (i.e., the net flux)
that is applied. 
Turbulent mixing of precursor gases is handled by CLUBB, i.e., inside
box 7 in the schematic in Fig.~\ref{fig:flowchart_EAM}.

\subsection{A simple revision to the original coupling scheme}
\label{sec:revised_coupling}

The discussions above motivates an alternate numerical scheme that couples the surface emissions
of dust (as well as other aerosol species) more tightly with the turbulent mixing. 
From a mathematical perspective, the ideal approach would be to provide the surface emission fluxes 
as a boundary condition to the equations 
of turbulent mixing so that the two processes (emissions and mixing) can be solved together numerically.
This approach would require a significant amount of code modifications and is deferred to future work.
Here we take a simpler and admittedly less optimal method 
that requires the least amount of code modifications, 
namely to move
the update of aerosol and precursor gas mixing ratios using {\sf cam\_in\%cflx}
from the original location to right before 
the cloud macrophysics-microphysics sub-cycles. 
In other words, box 3 in Fig.~\ref{fig:flowchart_EAM}
is moved to the location indicated by the pink box with dashed outline,
after box 6 (deep convection) and before box 7 (in which turbulent mixing is calculated).
The emission fluxes, all other parameterizations, and the
resolved dynamics are calculated at their original locations in the time loop.
For aerosols,
this simple modification still uses sequential splitting between emissions
and turbulence-mixing-activation-resuspension, but the dry removal processes are calculated 
before the surface emissions are applied, 
and the wet removal processes are calculated after 
turbulent mixing; hence neither the dry removal nor the wet removal
uses mixing ratios with spikes in the lowest model layer.

%% file: 4_impact_on_E3SM.tex
\section{Impact of the revised process coupling on aerosol climatology in EAMv1}
\label{sec:impact_on_E3SM}

We now compare EAMv1 simulations conducted with the original and revised coupling schemes.
The goal is twofold: 
(1) to verify the reasoning in the previous section about the features of the two schemes, and
(2) to evaluate the impacts of the revision on the simulated aerosol climatology at regional and global scales.

\subsection{Dust aerosols}
\label{sec:EAM_results_dust}

From the discussions in Sect.~\ref{sec:motivation_for_revision},
we expect the direct effect of the revised coupling scheme
to be a weakening of dry removal in grid cells and timesteps where dust emissions occur.
Based on the understanding of process interactions in EAM,
we can reason how the other aerosol processes can be affected.
In the revised scheme, turbulent mixing is the first aerosol process
calculated after surface emissions are applied.
Since the newly emitted particles have not gone through dry removal,
more (compared to the case in the original scheme) particles
are available for upward transport by turbulence.
After turbulent mixing, more aerosol particles can be expected 
in upper model layers in the source regions.
These particles can go through wet removal, 
or get advected out of the atmosphere column by resolved winds.
More transport from source to non-source regions
can increase aerosol load in the non-source regions
and consequently lead to more removal there.
These expectations are confirmed by the EAM results shown below.

%------ dust burden tendency maps, revised vs default EAMv1 -------------
\begin{figure*}[htbp]
\includegraphics[width=\textwidth]{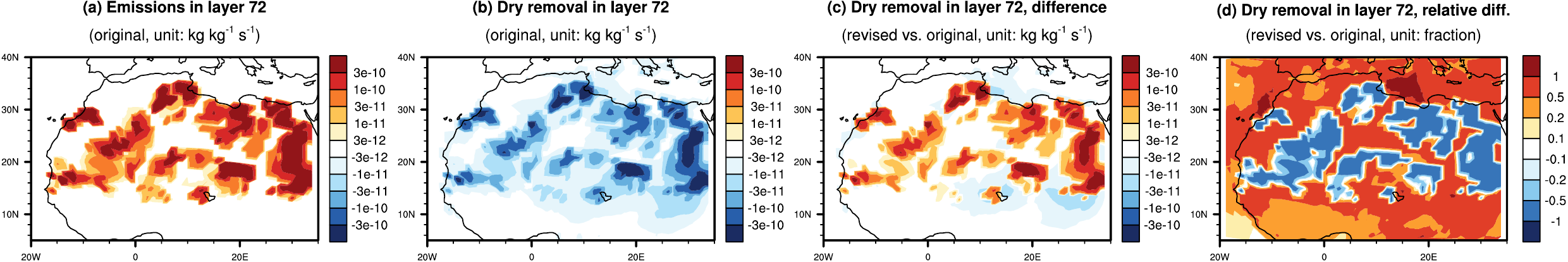}
\caption{\label{fig:dst_emis_dryrm_tend_lowest_layer_cmpr}
(a) and (b): Annual mean interstitial dust mass mixing ratio tendencies
(unit: kg~kg$^{-1}$~s$^{-1}$)
in the lowest model layer in simulation v1\_ori\_L72
caused by surface emissions and dry removal.
(c) and (d): The differences (unit: kg~kg$^{-1}$~s$^{-1}$) and
relative differences (unit: fraction) in dust dry removal 
between simulations using the revised and original process coupling schemes.
}
\end{figure*}

\begin{figure}[htbp]\vspace{5mm}
\includegraphics[width=0.4\textwidth]{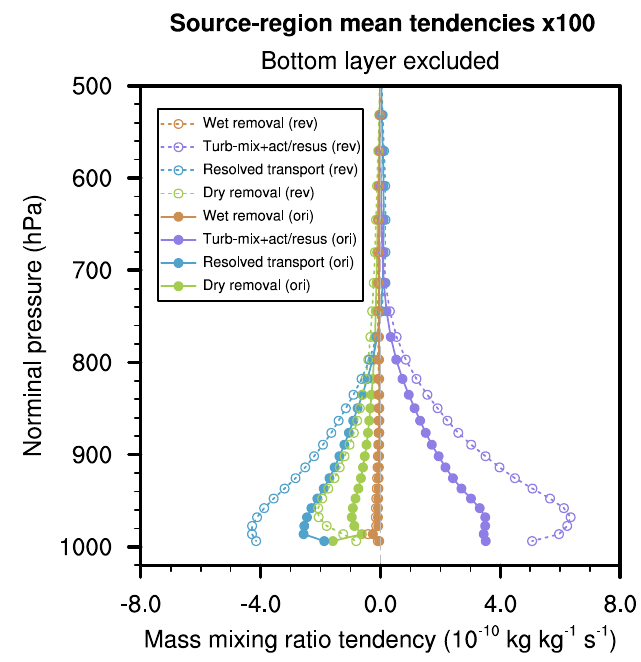}
\caption{\label{fig:source_region_ann_tend_profiles_rev_vs_ori}
Same as Fig.~\ref{fig:dust_sources_and_sinks}c 
but comparing results obtained with the original coupling scheme (filled circles)
and the revised coupling scheme (open circles).
Shown are annual mean results averaged over dust source regions located in
the magenta box in Fig.~\ref{fig:dust_sources_and_sinks}a.
}
\end{figure}

\begin{figure*}[htbp]
\includegraphics[width=0.96\textwidth]{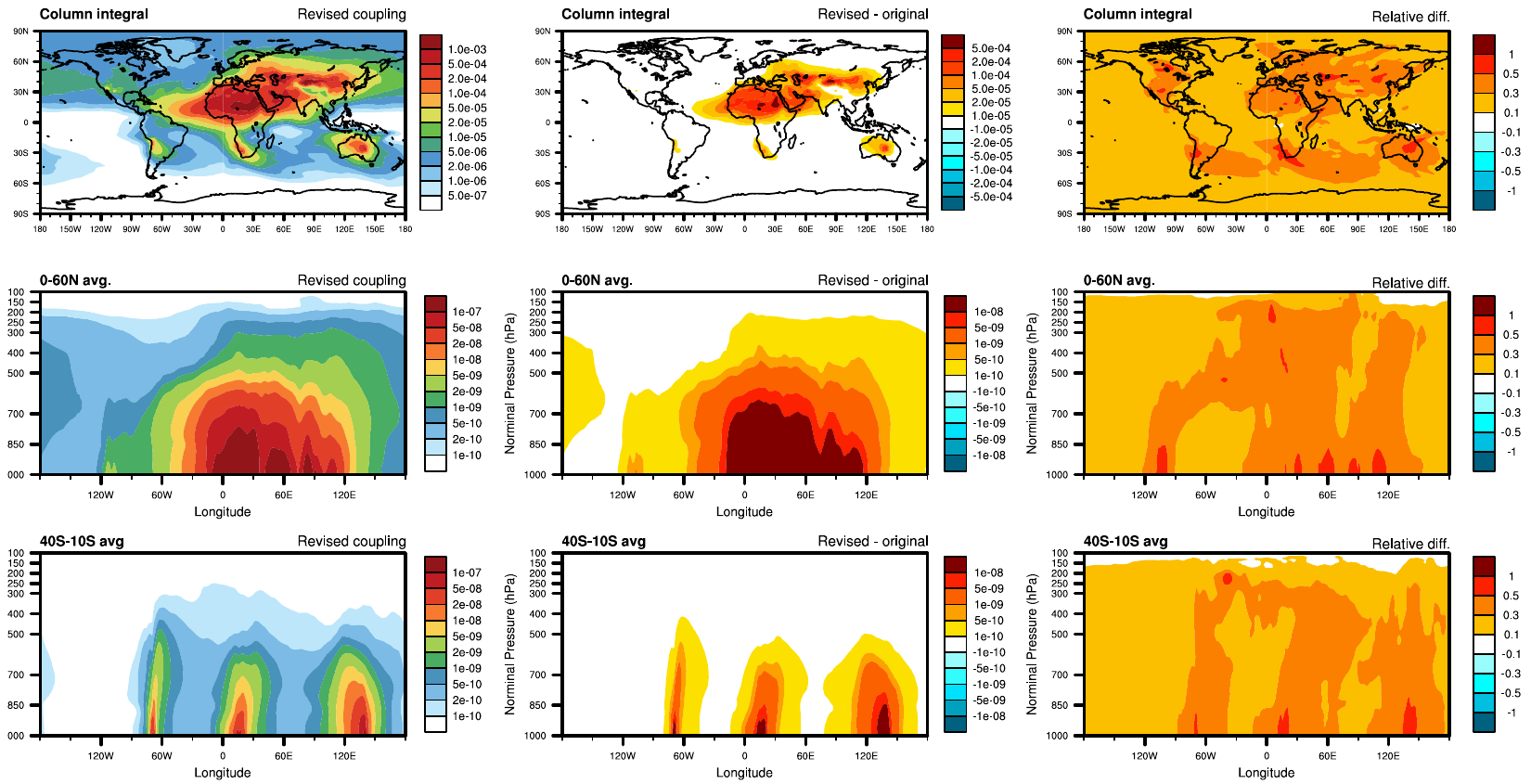}
\caption{\label{fig:dust_mass_rev_vs_ori}
Comparison of annual mean interstitial dust load simulated with the original or revised
process coupling scheme in EAMv1 using the L72 grid.
Top row: global distribution of vertically integrated dust burden
(unit: kg~m$^{-2}$).
Second and third rows: longitude-pressure cross-sections 
of interstitial dust mass mixing ratio (unit: kg~kg$^{-1}$) averaged over the latitude bands of
0$^\circ$--60$^\circ$N and 40$^\circ$S--10$^\circ$S, respectively.
Left column shows the L72 simulation using the revised coupling scheme
(v1\_rev\_L72).
Middle column shows the changes caused by the revised coupling
(i.e., v1\_rev\_L72 minus v1\_ori\_L72).
Right column shows the relative differences with respect to the original model.
}
\end{figure*}

The weakening of dry removal in the lowest model layer in dust source regions 
can be seen in Fig.~\ref{fig:dst_emis_dryrm_tend_lowest_layer_cmpr}
which shows annual mean results in North Africa as an example.
The emission-induced dust mixing ratio tendencies are shown in panel (a) to help identify 
locations with emissions.
The changes in dry removal tendencies caused by the revision of coupling 
show spatial patterns that closely match the emissions 
(Fig.~\ref{fig:dst_emis_dryrm_tend_lowest_layer_cmpr}c vs. \ref{fig:dst_emis_dryrm_tend_lowest_layer_cmpr}a). 
The relative differences shown in Fig.~\ref{fig:dst_emis_dryrm_tend_lowest_layer_cmpr}d 
suggest that 
decreases exceeding 50\% can be found in the majority of the North African dust source regions.

The process rate changes in dust source regions in model layers above the lowest 
can be seen from the tendencies profiles shown in Fig.~\ref{fig:source_region_ann_tend_profiles_rev_vs_ori}.
The figure is essentially the same as Fig.~\ref{fig:dust_sources_and_sinks}c 
but with the results from the revised coupling added as open circles.
Like in Fig.~\ref{fig:dust_sources_and_sinks}c, the tendencies shown here are annual mean values
averaged over dust source regions located in the magenta box in Fig.~\ref{fig:dust_sources_and_sinks}a.
The comparison in Fig.~\ref{fig:source_region_ann_tend_profiles_rev_vs_ori} shows 
that when the revised coupling is used,
large and systematic increases in magnitude are seen in turbulent mixing, resolved transport, and dry removal
in the lower troposphere above the lowest model layer. 
Between 900~hPa and 1000~hPa, the relative magnitude increases are on the order of 100\%.

The global impacts of the revised coupling can be seen in Fig.~\ref{fig:dust_mass_rev_vs_ori}
which shows the annual mean dust load in the atmosphere.
The first row corresponds to the interstitial dust burden integrated from the Earth's surface to model top.
The second and third rows are pressure-longitude cross-sections of 
dust mixing ratios meridionally averaged over
the two latitude bands of 0-60$^{\circ}$N and 40$^{\circ}$S-10$^{\circ}$S, respectively.
%The last row shows the pressure-latitude cross-section of the zonally averaged dust mixing ratios.
The burden or mixing ratio changes shown in the middle and right columns of the figure
indicate large and systematic increases in dust load, 
with relative increases of 30\% or higher within or near the source regions,
as well as typical increases between 10\% to 30\% in the remote regions 
(Fig.~\ref{fig:dust_mass_rev_vs_ori}, right column).

In Table~\ref{tab:dust_source_sink_rev_vs_ori}, 
the vertically integrated annual mean dust sources, sinks, burden,
and lifetime are shown as averages over all
source regions, over all regions without emissions, and over the entire globe.
The dust mass sinks attributed to activation
are from the parameterization of turbulent mixing and activation-resuspension,
noting that the mass-weighted column integral of the mixing ratio tendency caused by turbulent mixing vanishes.
For interstitial aerosols, the column-integrated wet removal rates indicate the
net effect of below-cloud wet removal (which converts interstitial aerosols to precipitation-borne aerosols)
and aerosol resuspension from evaporating precipitation (which converts precipitation-borne aerosols to interstitial aerosols).  
The numbers in the table further confirm that the revised coupling 
results in substantially weakened dry removal in dust source regions (36\% decrease),
stronger transport to non-source regions (56\% increase), 
and an overall (global mean) weakening of dry removal as well as strengthening of wet removal 
and activation.
The global mean interstitial dust burden increases by 39\%
and the global mean lifetime increases by 37\%.

\begin{table*}[htbp]
\vspace{5mm}
\caption{\label{tab:dust_source_sink_rev_vs_ori}
Impact of the revised process coupling on dust life cycle in EAMv1 simulations
conducted with the L72 vertical grid for the year 2010.
Shown here are the annual mean, vertically integrated sources, sinks, and burden of interstitial dust aerosol mass 
averaged over all grid columns with dust emission, all grid columns without dust emissions,
or the entire globe, as well as the relative changes caused by revising the coupling.
The lifetimes shown in the last row were calculated from the corresponding
values of mean burden and total source.
Units of the process rates, mass burden, and lifetime are indicated in the left most column.
Relative differences are shown as percentages.
}
\scalebox{0.9}{
\begin{tabular}{lrrrrrrrrrrrr}\tophline
\multirow{ 2}{*}{\bf Dust budget}  
                           &  \multicolumn{3}{c}{\bf Source-region mean} && \multicolumn{3}{c}{\bf Non-source-region mean} && \multicolumn{3}{c}{\bf Global mean} \\\cline{2-12}
                           &    Original &     Revised & Rel. diff. && Original & Revised & Rel. diff. && Original & Revised & Rel. diff. \\\middlehline
Dry removal        (Tg yr$^{-1}$) &  $-$69069.1 &  $-$44272.6 &  $-$36\% &&   $-$684.2 &   $-$1216.6 & $+$78\% &&  $-$2982.6 &   $-$2663.8 &  $-$11\% \\
Activation         (Tg yr$^{-1}$) 
                                  &   $-$3037.3 &   $-$4612.5 &  $+$52\% &&   $-$296.1 &    $-$405.3 & $+$37\% &&   $-$388.3 &    $-$546.8 &  $+$41\% \\
Wet removal         (Tg yr$^{-1}$) 
                                  &   $-$3146.7 &   $-$4971.4 &  $+$58\% &&   $-$431.0 &    $-$585.1 & $+$36\% &&   $-$522.3 &    $-$732.5 &  $+$40\% \\
Resolved transport (Tg yr$^{-1}$) &  $-$40808.4 &   $-$63825.1 & $+$56\% &&   $+$1411.3 &  $+$2207.0 & $+$56\% &&          - &           - &        - \\
\middlehline
Burden (Tg)                       &       213.6 &       311.9 &  $+$46\% &&       13.2 &        17.8 & $+$35\% &&       20.0 &        27.7 &  $+$39\% \\
Lifetime (day)                    &         0.7 &         1.0 &  $+$44\% &&        3.4 &         2.9 & $-$14\% &&        1.9 &         2.6 &  $+$37\% \\                          %
\bottomhline
\end{tabular}
}%scalebox
\end{table*}

\subsection{Other species}
\label{sec:EAM_results_other_species}

Sect.~\ref{sec:coupling_scheme_in_default_EAMv1} has explained that in 
the original EAMv1, surface emissions of aerosols 
are converted to mixing ratio tendencies and applied to the corresponding tracers 
at the location marked as 3 in the schematic in Fig.~\ref{fig:flowchart_EAM}.
The revision of process coupling evaluated in this paper moves
this block of calculations to the location between 6 and 7 in 
Fig.~\ref{fig:flowchart_EAM} (see pink box with dashed outline), meaning that 
not only dust but all aerosol species that have surface emissions
are affected by the code modification.
The precursor gases with non-zero net surface fluxes
(emission minus dry deposition) are affected as well.

In Sect.~\ref{sec:doc_emissions}, it has been explained that
5 out of the 7 aerosol species in MAM4, namely dust, sea salt, 
MOA, BC, and POA, have large portions (50\% or more) of their
sources coming from surface emissions. The other two species,
sulfate and SOA, are predominantly or 100\% produced in the atmosphere,
as can be seen in the attribution of mass sources shown in Table~\ref{tab:emissions}.
Therefore, we expect the revised coupling to have significant impacts on the first 5 aerosol species
mentioned above but negligible direct impacts on sulfate and SOA.

In Table~\ref{tab:source_sink_rev_vs_ori_all_species}, we show 
the vertically integrated and annually averaged dry removal rates
for regions with or without surface emissions as well as for the entire globe,
for the 5 aerosol species with substantial surface sources.
The relative differences listed in the table
reveal consistent results among the different species, namely
dry removal becomes weaker in regions with surface emissions
but stronger in regions without surface emissions.
The global averages, which are dominated by the results in the source regions,
show a general trend of weaker dry removal when the revised coupling is used.

\begin{table*}[htbp]
\vspace{5mm}
\caption{\label{tab:source_sink_rev_vs_ori_all_species}
Similar to Table~\ref{tab:dust_source_sink_rev_vs_ori} but showing the
impact of the revised process coupling 
on dry removal of aerosol species that have substantial surface emissions.
Here, source region and non-source region refer to geographical locations with or without
surface emissions of the corresponding aerosol species.
Dry removal rates are shown in Tg~yr$^{-1}$.
Relative differences are shown in percentages.
}
\scalebox{0.95}{
\begin{tabular}{lrrrrrrrrrrrrrr}\tophline
\multirow{ 2}{*}{\bf Dry removal}
&  \multicolumn{3}{c}{\bf Source-region mean} && \multicolumn{3}{c}{\bf Non-source-region mean} && \multicolumn{3}{c}{\bf Global mean} \\\cline{2-12}
         & Original & Revised & Rel. diff. && Original & Revised & Rel. diff. && Original & Revised & Rel. diff. \\\middlehline
Dust     &  $-$69069.1 &  $-$44272.6 &$-$36\% &&   $-$684.2 &   $-$1216.6 & $+$78\% &&  $-$2982.6 &   $-$2663.8 & $-$11\% \\
Sea salt &   $-$2927.3 &   $-$1762.9 &$-$40\% &&    $-$66.3 &    $-$114.5 & $+$73\% &&  $-$2198.9 &   $-$1343.3 & $-$39\% \\
MOA      &      $-$9.5 &      $-$5.6 &$-$41\% &&     $-$0.9 &      $-$1.1 & $+$17\% &&     $-$6.8 &      $-$4.2 & $-$39\% \\
BC       &      $-$3.7 &      $-$2.9 &$-$21\% &&     $-$0.0 &      $-$0.0 & $+$19\% &&     $-$3.6 &      $-$2.8 & $-$21\% \\
POA      &     $-$16.7 &     $-$14.1 &$-$16\% &&     $-$0.3 &      $-$0.3 &  $+$7\% &&    $-$16.1 &     $-$13.6 & $-$16\% \\
\bottomhline
\end{tabular}
}%scalebox
\end{table*}

\begin{table*}[htbp]\vspace{5mm}
\caption{\label{tab:burden_lifetime_rev_vs_ori_all_species}
Similar to Table~\ref{tab:source_sink_rev_vs_ori_all_species}
but showing the impact of the
revised process coupling on the global mean burden (unit: Tg) and
lifetime (unit: day) of different aerosol species.
Relative differences are shown in percentages.
}
\scalebox{0.95}{
\begin{tabular}{lrrrrrrrrrrrrrr}\tophline
\multirow{ 2}{*}{\bf Burden and lifetime}
&  \multicolumn{3}{c}{\bf Global mean burden} && \multicolumn{3}{c}{\bf Global mean lifetime} \\\cline{2-8}
            & Original & Revised & Rel. diff. && Original & Revised & Rel. diff. \\\middlehline
Dust        &     20.0 &    27.7 &    $+$39\% &&    1.91  &    2.62 &   $+$37\%  \\
Sea salt    &     4.30 &    6.54 &    $+$52\% &&    0.53  &    0.81 &   $+$53\%  \\
MOA         &    0.078 &   0.092 &    $+$18\% &&    1.43  &    1.68 &   $+$17\% &&  \\
BC          &    0.16  &   0.18  &    $+$13\% &&    6.18  &    6.97 &   $+$13\% &&  \\
POA         &    0.94  &   1.02  &     $+$9\% &&    7.24  &    7.80 &    $+$8\% &&  \\
\bottomhline
\end{tabular}
}%scalebox
\end{table*}

%------ zonal mean mass mixing ratio of 5 species, revised vs default EAMv1 -------------
\begin{figure*}[htbp]
\includegraphics[width=0.9\textwidth]{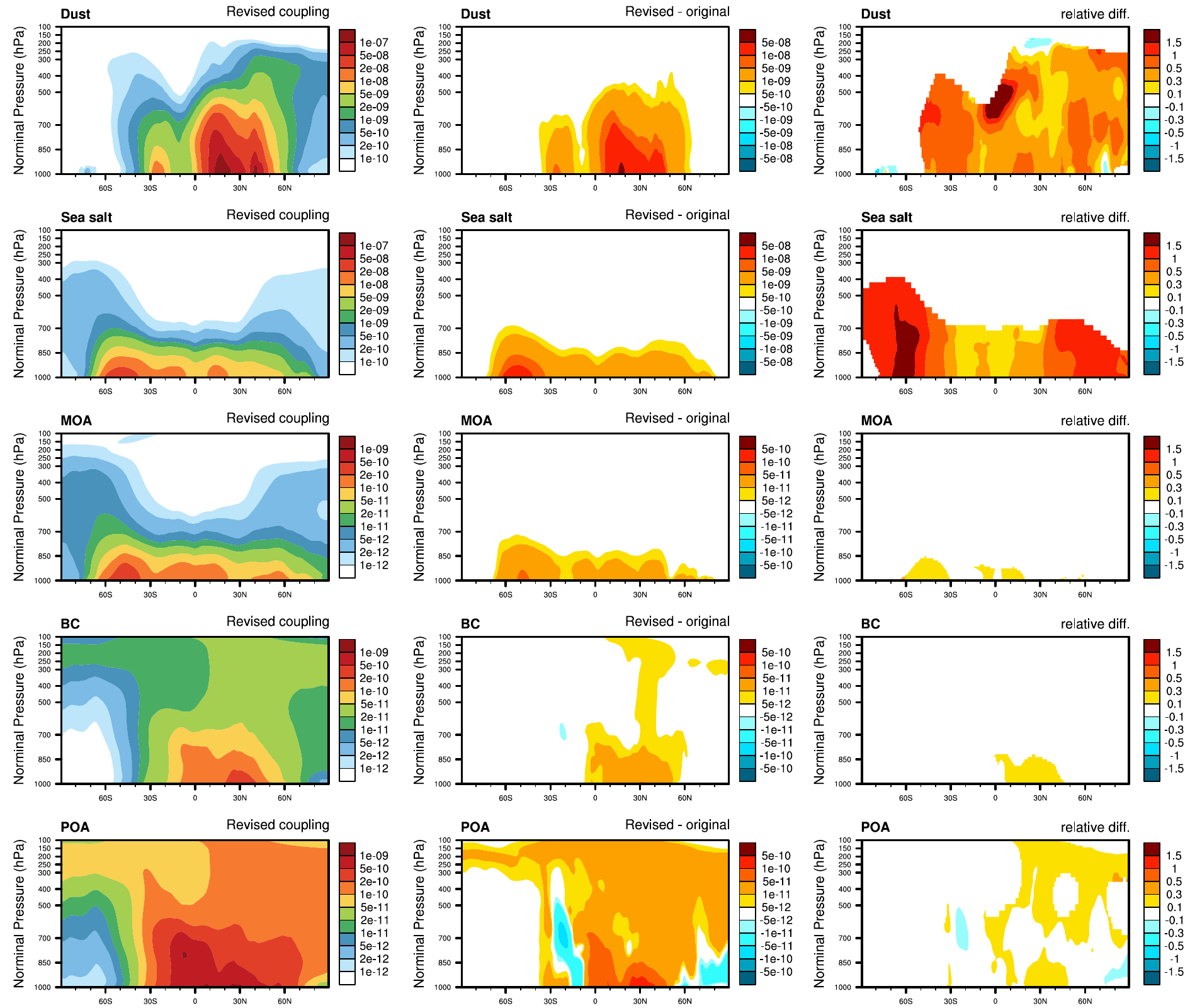} 
\caption{\label{fig:zonal_mean_mixing_ratio_rev_vs_ori_5_aerosols}
Comparison of the zonal and annual mean interstitial aerosol mass mixing ratios (unit: kg~kg$^{-1}$)
simulated with EAMv1 using the original or revised process coupling and with the L72 vertical grid.
The different rows correspond to the 5 aerosol species listed in Table~\ref{tab:emissions_by_mode}. 
Left column shows the L72 simulation using the revised coupling scheme.
Middle column shows the differences caused by the revision in process coupling.
Right column shows the relative differences with respect to the original model.
For the reference differences plots, the locations with mean mixing ratios
lower than the first contour level in the left panel are masked out.
}
\end{figure*}

It is worth noting that among those 5 aerosol species that have substantial surface emissions,
dust and sea salt emissions introduce particles primarily 
to MAM4's coarse mode
(where the mode's geometric mean diameter range is 1~$\mu$m~$\leqslant D_p\leqslant$~4~$\mu$m),
while MOA, BC, and POA emissions add particles mostly to MAM4's
primary carbon mode (10~nm~$\leqslant D_p\leqslant$~100~nm)
and accumulation mode (53.5~nm~$\leqslant D_p\leqslant$~440~nm).
This can be seen from the emission attribution 
shown in Table~\ref{tab:emissions_by_mode}. 
Although there are various mechanisms for different aerosol species to mix
and for aerosol particles to grow in the atmosphere,
MOA, BC, and POA are predominantly found in aerosol particles of submicron diameters.
Since smaller particles are less susceptible to dry removal,
we expect to see smaller impacts of the revised coupling on the
atmospheric load of MOA, BC, and POA than on dust and sea salt.
This reasoning is confirmed by 
the global annual mean mass burdens shown in Table~\ref{tab:burden_lifetime_rev_vs_ori_all_species}
and the zonally and annually averaged mass mixing ratios shown in
Fig.~\ref{fig:zonal_mean_mixing_ratio_rev_vs_ori_5_aerosols}.
The global mean burdens increase by 39\% and 52\% for dust and sea salt, respectively,
but less than 20\% for MOA, BC, and POA.
The relative changes in zonal mean mixing ratio shown in the rightmost column
in Fig.~\ref{fig:zonal_mean_mixing_ratio_rev_vs_ori_5_aerosols}
further confirm the large impacts on dust and sea salt
and the smaller impacts on the predominantly submicron species (MOA, BC, POA).
As for sulfate, SOA, and precursor gases in EAMv1,
the changes in zonal and annual mean 
mixing ratios are typically less than 10\% and hence not shown.

\subsection{Sensitivity to bottom layer thickness}
\label{sec:EAM_results_resolution_sensitivity}

In Sect.~\ref{sec:original_coupling_weakness}, we attributed
the strong vertical resolution sensitivity of 
dust dry removal reported in \citet{FengY_2022_JAMES_dust_in_EAMv1}
to the original sequential splitting scheme that calculates dry removal after
surface emissions and before turbulent mixing.
The revised coupling scheme allows the emitted particles to be
vertically mixed by turbulence before other processes are calculated.
Since turbulent mixing is very efficient in reducing the vertical gradients
and resulting in similar mixing ratios in model layers near the Earth's surface
(Fig.~\ref{fig:dust_state_box_plot_source}), we expect the
simulated dry removal rates to be less sensitive to 
the thickness of the lowest model layer when the revised coupling is used, 
and we expect the same qualitative results to be seen in all of the 
5 aerosol species that have significant surface emissions.
These expectations are confirmed by the results shown in Table~\ref{tab:dry_removal_L72_L30}.

Furthermore, the global and annual mean aerosol lifetimes shown in 
Table~\ref{tab:lifetimes} indicate that when the revised coupling is used,
the relative differences between results from the L30 and L72 simulations
are reduced to less than 10\% in contrast to 32\% for dust and 47\% for sea salt in the original EAMv1.

\begin{table*}[htbp]
\vspace{5mm}
\caption{\label{tab:dry_removal_L72_L30}
Sensitivity of the annual (year-2010) mean
source-region mean aerosol dry removal rates (unit: Tg~yr$^{-1}$)
to vertical resolution when using the original or revised process coupling in EAMv1.
Results are shown for the 5 aerosol species in MAM4 that have significant surface emissions.
}
\scalebox{0.95}{
\begin{tabular}{lrrrrrrrrrrrrrr}\tophline
\multirow{ 2}{*}{\bf Source-region mean dry removal}
& \multicolumn{3}{c}{\bf Original coupling} && \multicolumn{3}{c}{\bf Revised coupling} \\\cline{2-8}
         & L72         &    L30      & Rel. diff. &&    L72 &         L30 & Rel. diff.\\\middlehline
Dust     &  $-$69069.1 &  $-$41026.7 &    $-$41\% && $-$44272.6 &  $-$36915.4 &  $-$17\%  \\
Sea salt &   $-$2927.3 &   $-$1947.8 &    $-$33\% &&  $-$1762.9 &   $-$1689.7 &   $-$4\% \\
MOA      &      $-$9.5 &      $-$5.6 &    $-$41\% &&     $-$5.6 &      $-$5.0 &  $-$11\%\\
BC       &      $-$3.7 &      $-$2.9 &    $-$23\% &&     $-$2.9 &      $-$2.7 &   $-$7\%\\
POA      &     $-$16.7 &     $-$13.5 &    $-$19\% &&    $-$14.1 &     $-$13.0 &   $-$8\%\\
\bottomhline
\end{tabular}
}%scalebox
\end{table*}

\begin{table*}[htbp]
\vspace{5mm}
\caption{\label{tab:lifetimes}
Sensitivity of the global and annual (year-2010)
mean aerosol lifetime (unit: day)
to vertical resolution when using the original or revised process coupling in EAMv1.
Results are shown for the 5 aerosol species in MAM4 that have significant surface emissions.
}
\scalebox{0.95}{
\begin{tabular}{lrrrrrrrrrrrrrr}\tophline
\multirow{ 2}{*}{\bf Global mean lifetime}
         & \multicolumn{3}{c}{\bf Original coupling} && \multicolumn{3}{c}{\bf Revised coupling} \\\cline{2-8}
         & L72  & L30   & Rel. diff. && L72  & L30  & Rel. diff.\\\middlehline
Dust     & 1.91 & 2.52  &    $+$32\% && 2.62 & 2.68 &     $+$2\%\\
Sea salt & 0.53 & 0.78  &    $+$47\% && 0.81 & 0.86 &     $+$6\%\\
MOA      & 1.43 & 1.71  &    $+$20\% && 1.68 & 1.77 &     $+$4\%\\
BC       & 6.18 & 6.74  &     $+$9\% && 6.97 & 6.87 &     $-$1\%\\
POA      & 7.24 & 7.50  &     $+$4\% && 7.80 & 7.59 &     $-$3\%\\
\bottomhline
\end{tabular}
}%scalebox
\end{table*}

%% file: 5_conclusions.tex
\conclusions[Summary, conclusions, and outlook]
\label{sec:conclusions}

The earlier work by \citet{FengY_2022_JAMES_dust_in_EAMv1} pointed out that various aspects of
the dust aerosol life cycle simulated by EAMv1, in particular the dry removal fluxes and lifetime,
were sensitive to the use of 72 versus 30 layers for the vertical grid.
In this paper, we investigated the resolution sensitivity by
carrying out detailed budget analyses for the interstitial dust mass mixing ratio
and by tracking the evolution of the mixing ratio within the model's time integration cycle.
Dust emissions in the real world typically occur in turbulent
ambient conditions, hence the emitted particles can be
efficiently distributed to a significant depth of the atmosphere column.
In EAMv1, the numerical coupling method treats surface emissions as a separate process
which adds aerosol particles to the lowest model layer, while dry removal and turbulent mixing
are calculated after the emissions are applied, using the sequential splitting method.
This ordering of processes
results in high, unrealistic temporary values of aerosol mixing ratios to be
provided as input to the parameterization of dry removal, 
causing large numerical errors in the simulated dry removal rates.
The situation gets worse when the model's bottom layer is thinner,
as the same emission fluxes will result in higher temporary mixing ratios in
the bottom layer when emissions are applied in isolation.

Based on this reasoning, we proposed a simple revision to the numerical process coupling in EAMv1,
namely to move the application of emissions to the location right before turbulent mixing
in the model's time integration loop.
The revision allows the newly emitted aerosol particles to be vertically distributed by turbulence
before experiencing other processes considered in the model, hence giving a numerical coupling scheme
with closer resemblance to the process interactions in the real world.

The revised coupling scheme was implemented in EAMv1, and wind-nudged simulations were conducted
for the year 2010 using 72 or 30 grid layers with 1$^\circ$ horizontal grid spacing.
As expected, the revision substantially 
weakened dry removal and strengthened vertical mixing of dust in its source regions,
strengthened the large-scale transport from source to non-source regions,
strengthened dry removal outside the source regions, and strengthened
activation and wet removal globally.
When using 72 grid layers without retuning of uncertain parameters,
the revised process coupling was found to cause a 39\% increase in the global annual mean dust burden
and an increase of dust lifetime from 1.9 days to 2.6 days.

The revised process coupling was implemented for all aerosol species in EAMv1 
and was found to cause the same qualitative changes also
for other aerosol species that have substantial surface emissions, 
i.e., sea salt, MOA, BC, and POA.
Quantitatively, the changes in global mean burden and lifetime were
considerably smaller for the predominantly submicron species
(MOA, BC, POA), with burden and lifetime increases less than 20\%, whereas 
the increases in dust and sea salt burden and lifetime exceeded 30\% and 50\%, respectively
(Table~\ref{tab:burden_lifetime_rev_vs_ori_all_species}).
The impact on sulfate, SOA, and precursor gases were found to be very small.

%-------------
% future work
%-------------

The revised process coupling proposed in this study allowed us to
use a minimal amount of code changes to
obtain significant improvements in the EAMv1 results.
However, both the surface emissions and dry removal of aerosols are 
still sequentially split from turbulent mixing
using relatively long coupling timesteps of 30 minutes.
This is likely a significant contributor to the remaining
17\% difference in the annual mean source-region mean dust dry removal rate
between the L72 and L30 simulations when the revised coupling is used
(Table~\ref{tab:dry_removal_L72_L30}).
In the next steps, it will be worth exploring additional revisions
to further reduce the numerical errors caused by process splitting.
For example, since emissions in the real world are typically
accompanied by turbulent mixing, we plan to implement and evaluate
a scheme that combines the time integration of emissions and turbulent mixing.
Since turbulent dry deposition is also closely related to
turbulent mixing, it is worth considering combining these two processes as well.
In contrast, it is less obvious which coupling methods 
will be cost-effective in reducing the splitting errors related to gravitational settling.
Different options will be explored in future work.
We do not expect that vertical resolution sensitivities
in the simulated aerosol life cycles will be completely eliminated
merely by further revising the numerical coupling of aerosol processes discussed in this paper,
because discretization errors in the individual parameterizations,
as well as discretization errors and numerical coupling errors
associated with other processes in EAM (e.g., clouds), can also cause vertical resolution sensitivities,
and some of the sensitivities can be advantageous as they demonstrate the benefits of using higher resolutions.
Nevertheless, further work on numerical coupling will likely
make useful contributions to the goal of
reducing numerical errors in EAM simulations.

%% file: appendix_E3SM_figures.tex
\section{Additional figures}    %% Appendix A

\begin{figure*}[htbp]
\vspace{5mm}
\includegraphics[width=\textwidth]{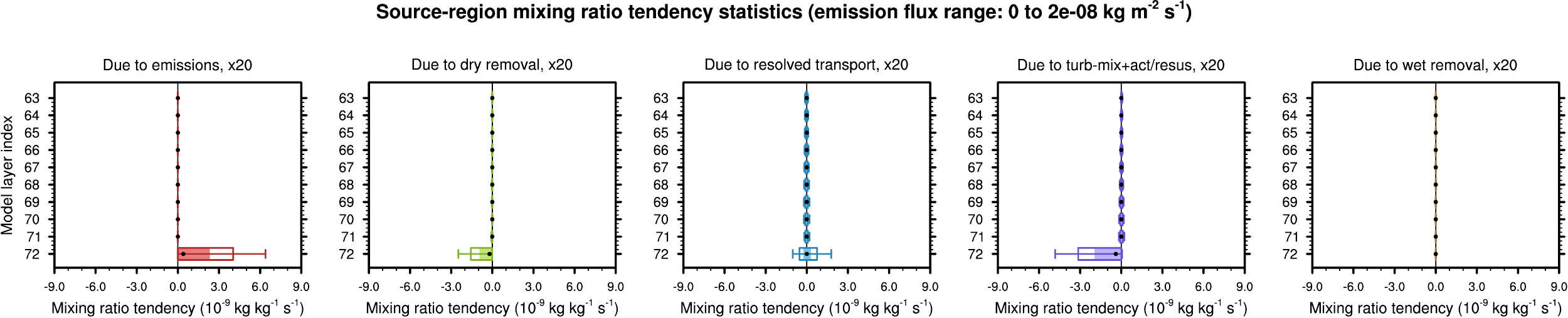}\vspace{4mm}
\includegraphics[width=\textwidth]{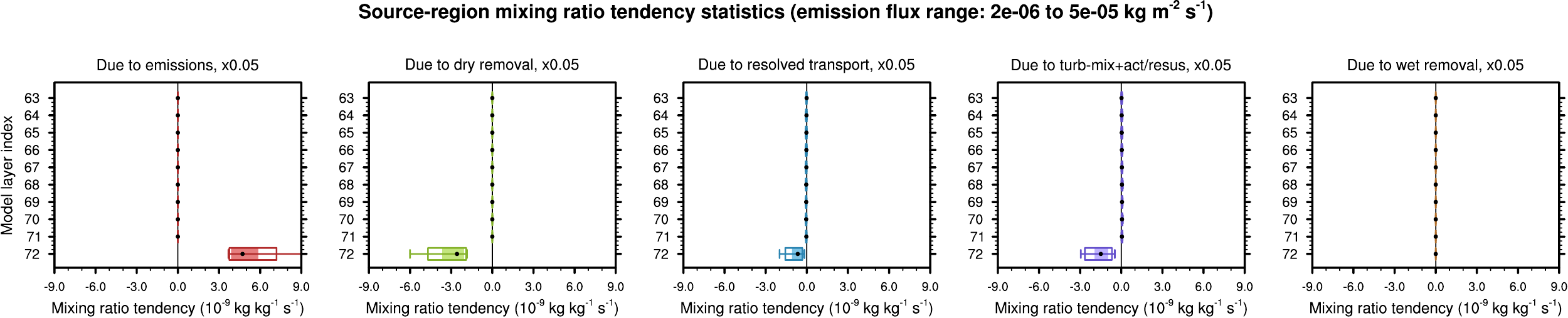}
\caption{
As in Fig.~\ref{fig:dust_tend_box_plot_source} but with the upper row showing results
for the grid columns and timesteps corresponding to the left portion of the histogram in Fig.~\ref{fig:emission_pdf};
the lower row shows results for the right portion of that same histogram in Fig.~\ref{fig:emission_pdf}.
The mixing ratio tendencies in the upper and lower rows have been multiplied 
by 20 and 0.05, respectively, in order to be plotted with the same x-axis range 
as in Fig.~\ref{fig:dust_tend_box_plot_source}.
Results for the source-vicinity and remote regions are shown 
in Fig.~\ref{fig:dust_tend_box_plot_vicinity_and_remote}.
\label{fig:dust_tend_box_plot_source_2}}
\end{figure*}

\begin{figure*}[htbp]
\includegraphics[width=\textwidth]{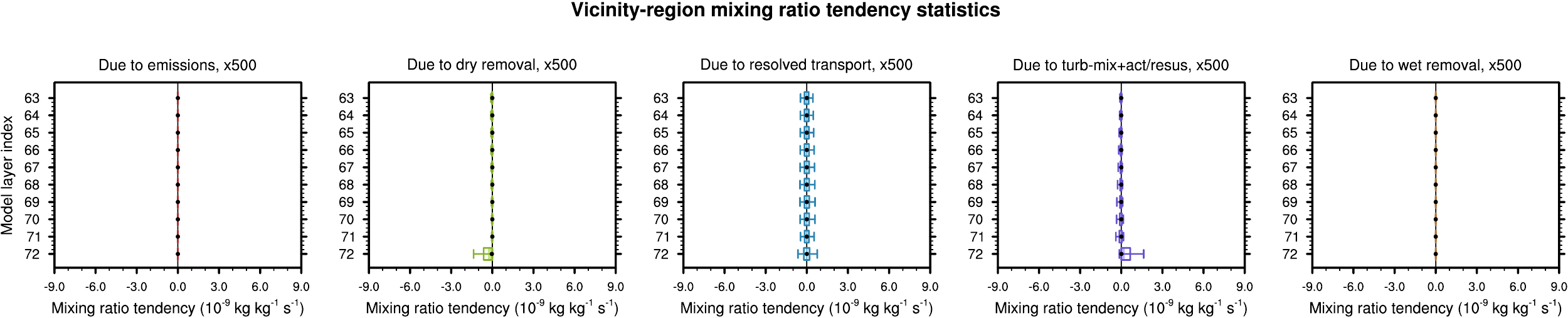}\vspace{5mm}
\includegraphics[width=\textwidth]{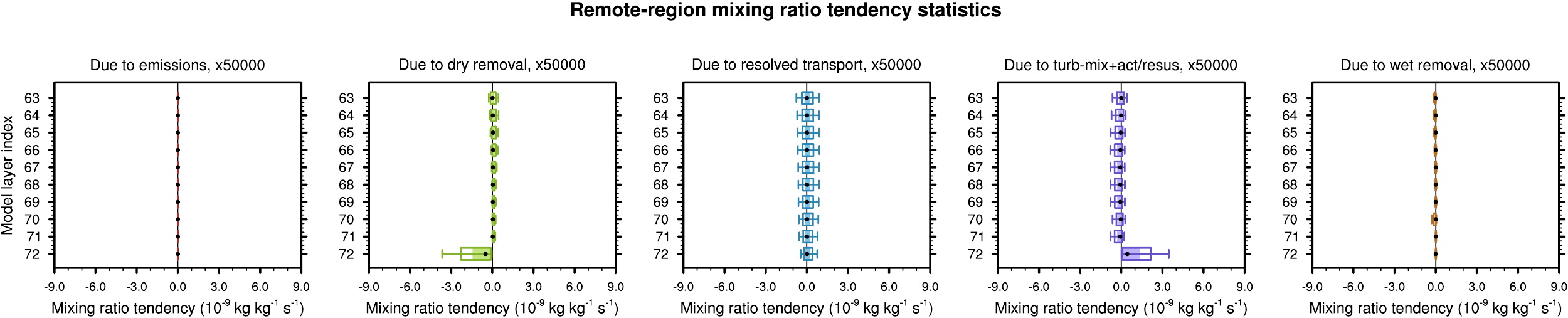}
\caption{
As in Figs.~\ref{fig:dust_tend_box_plot_source} and \ref{fig:dust_tend_box_plot_source_2}
but showing results 
in the source-vicinity regions (upper row) and 
remote regions (lower row). 
The source-vicinity regions are defined as grid columns and timesteps with zero dust emission 
in the 90-day period of 2010-01-29 to 2010-04-28
in the magenta box marked in Fig.~\ref{fig:dust_sources_and_sinks}a.
The remote regions are defined as the brown box in Fig.~\ref{fig:dust_sources_and_sinks}a.
The statistical distributions were derived from 6-hourly instantaneous output
from the default EAMv1 (i.e., simulation v1\_ori\_L72 in Table~\ref{tab:simulations}).
The mixing ratio tendencies in the upper and lower rows have been multiplied 
by 500 and 50,000, respectively, in order to be plotted with the same x-axis range 
as in Fig.~\ref{fig:dust_tend_box_plot_source}.
}
\label{fig:dust_tend_box_plot_vicinity_and_remote}
\end{figure*}

\begin{figure*}[htbp]
\includegraphics[width=\textwidth]{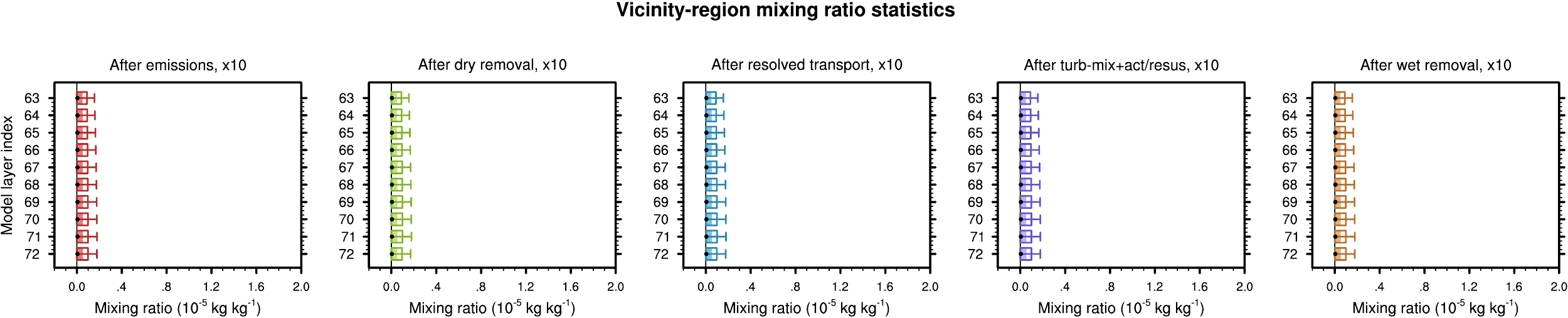}\vspace{5mm}
\includegraphics[width=\textwidth]{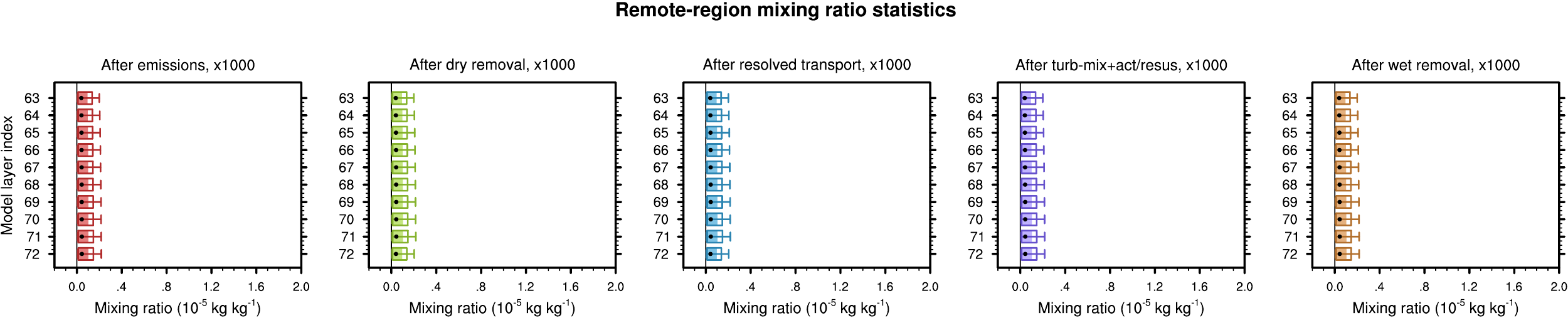}
\caption{
As in Fig.~\ref{fig:dust_state_box_plot_source} but showing  
statistical distributions of instantaneous dust mass mixing ratios 
in the source-vicinity regions (upper row) and 
remote regions (lower row). 
The mixing ratios in the upper and lower rows have been multiplied 
by 10 and 1,000, respectively, in order to be plotted with the same x-axis range 
as in Fig.~\ref{fig:dust_state_box_plot_source}.
}
\label{fig:dust_state_box_plot_vicinity_and_remote}
\end{figure*}

Figures~\ref{fig:dust_tend_box_plot_source_2} and \ref{fig:dust_tend_box_plot_vicinity_and_remote}
supplement Fig.~\ref{fig:dust_tend_box_plot_source}
by showing statistical distributions of interstitial dust mass mixing ratio
tendencies in dust source regions with weak or very strong emissions (Fig.~\ref{fig:dust_tend_box_plot_source_2})
and in regions without dust emissions 
(i.e., source-vicinity and remote regions, Fig.~\ref{fig:dust_tend_box_plot_vicinity_and_remote}).
Figure~\ref{fig:dust_state_box_plot_vicinity_and_remote} 
supplements Fig.~\ref{fig:dust_state_box_plot_source} by showing  
statistical distributions of interstitial dust mass mixing ratios 
in regions without dust emissions
(i.e., source-vicinity and remote regions).

%%------ dust burden tendency maps, revised vs default EAMv1 -------------
%\begin{figure*}[htbp]
%\includegraphics[width=0.9\textwidth]{dust_burden_tend_cmpr.pdf}
%\caption{\label{fig:dust_burden_tend_global}
%Comparison of the annual mean interstitial dust burden tendencies (unit: kg~m$^{-2}$~s$^{-1}$)
%simulated with the original or revised EAMv1. 
%The different rows correspond to different source and sink terms. 
%(Note that the two dry removal mechanisms, i.e., gravitational settling and turbulent dry deposition, are shown separately in this figure.)
%Left column shows results from the revised model.
%Middle column shows the difference between the revised model and the default EAMv1.
%Right column shows the relative difference with respect to the default model.
%}
%\end{figure*}